\documentclass[preprint]{elsarticle}

\usepackage{lineno,hyperref}
\usepackage{amsmath}
\usepackage{amssymb}
\usepackage{amsfonts}
\usepackage{graphicx,subcaption,float}
\usepackage{epstopdf}
\usepackage{tikz}
\usepackage{wasysym}

\usepackage{slashed}

\usepackage{geometry}
\modulolinenumbers[1]

\makeatletter
\def\ps@pprintTitle{%
  \let\@oddhead\@empty
  \let\@evenhead\@empty
  \def\@oddfoot{\reset@font\hfil\thepage\hfil}
  \let\@evenfoot\@oddfoot
}
\makeatother

\biboptions{sort&compress}

\begin{document}

\begin{frontmatter}

\title{Matching to Higgs-Compositeness and Renormalization of the Higgs-Electroweak Chiral Lagrangian extended by a Scalar Singlet}

\author[LMU]{Andreas Lindner}
\ead{and.lindner@physik.uni-muenchen.de}
\author[WWU]{Khoirul Faiq Muzakka}
\ead{khoirul.muzakka@uni-muenster.de}

\address[LMU]{Arnold Sommerfeld Center for Theoretical Physics, Ludwig-Maximilians-Universit\"at M\"unchen\\Theresienstr. 37, D-80333 M\"unchen, Germany}
\address[WWU]{Institut f{ü}r Theoretische Physik, Westf{ä}lische Wilhelms-Universit{ä}t M{ü}nster, Wilhelm-Klemm-Stra{ß}e 9, D-48149 M{ü}nster, Germany}

\begin{abstract}
We match the electroweak chiral Lagrangian with two singlet scalars to the next-to-minimal composite Higgs model with $ SO(6)/SO(5) $ coset structure and extract the scalar divergences to one loop.
Assuming the additional scalar to be heavy, we integrate it out and perform a matching to the well-established electroweak chiral Lagrangian with one light Higgs.

\end{abstract}


\end{frontmatter}
\tableofcontents{}

\section{Introduction}
The Higgs sector within the Standard Model is still mysterious even after the discovery of a "Higgs-like" particle.
Open questions regarding naturalness of the Higgs mass, and form and origin of the Higgs potential remain \cite{futurepp,Agrawaletal2020}.
At the same time these are open opportunities for further investigations.
The formalism of the electroweak chiral Lagrangian with a Higgs \cite{longhitano1980,longhitano1981,herrero1,herrero2,appelquist1980,feruglio} is considered the most general approach to open-mindedly take into account deviations from the Standard Model \cite{higgscoupling,fithiggs}.
The electroweak chiral Lagrangian was extended by two and then N scalar fields in our aforementioned work \cite{lindnermuzakka2022}.
A second scalar particle in addition to the observed one has been extensively discussed in, e.g., \cite{Redi,chala,qi2021effective}.
An additional scalar field can serve as a dark matter candidate \cite{dm} and lead to new CP violating processes required to explain the matter-antimatter asymmetry \cite{EWPhT,baryogenesis,chala}.
It brings a lot of new physics and widens the narrow corset of parameter space which constrains the electroweak chiral Lagrangian with only one scalar that necessarily has to be identified with the observed one.

Composite Higgs models (CHMs) \cite{kaplan,kaplan1,custodialsu2,dugan,Bellazzini} provide logical ideas of underlying extensions to electroweak theories, first proposed in the context of unified composite models of all fundamental particles and forces \cite{Terazawa1977,Terazawa1980}.
The composite Higgs model with one Higgs scalar has been worked on by many authors.
The minimal custodial CHM has the coset structure $ SO(5)/SO(4) $ and has been extensively studied, e.g., in \cite{minimal,so5,Wulzer}.
A concise review is given in \cite{contino} and the phenomenology was particularly emphasized in \cite{niehoff}, to name just a few.
A matching of the minimal model to the electroweak chiral Lagrangian with a light Higgs was performed in \cite{Krause} for the fermions in the spinorial \textbf{4} representation of $ SO(5) $.
This representation is, from our perspective today, unsuited for some reasons.
It does not contain the left-handed fermion doublets in a custodially symmetric representation, thus leading to unprotected $ Zq_Lq_L $ couplings \cite{niehoff,Wulzer}.
The fermion embedding is a decisive decision for the building of the model since the Goldstone potential is heavily dependent on it.
We are therefore choosing the next-to-minimal composite Higgs model with the coset $ SO(6)/SO(5) $ and the fermions in the fundamental \textbf{6} representation, short NMCHM6.
The model has originally been introduced in \cite{gripaios}.
There are two Higgs scalars in this model.
Phenomenological consequences and scans have been studied for $ f \lesssim 1TeV $ in \cite{Redi,niehoff}.
Interestingly, it is the minimal model permitting a fermionic UV completion, rendering it a successful and the most economic CHM \cite{fermionicUV,fundamentalcompositeness,Bizot,compositeUV,completecompUV}.

Composite Higgs models solve the hierarchy problem in a very intuitive and realistic way.
On the fly, the Higgs potential is generated dynamically and a mechanism explaining the Yukawa hierarchy is provided.
A very important aspect is the embedding of the quarks in multiplets of the full global group $ G $, as stated above.
This is the symmetry group under which the interactions of elementary and composite sector originate at the scale $ \Lambda_{\text{comp}} $, where strong resonance particles are supposed to live and mediate interactions.
That scale is assumed to be far above the scale of the spontaneous breaking $ G \rightarrow H $.
Varying representations give rise to very different interactions.
Interactions of the gauge sector with the strong sector, on the other hand, are determined by the coset $ G/H $ alone.

\subsection{Outline}

In Section \ref{unitarization} we address the problem of unitarization in composite Higgs models and Higgs effective theory and recap the concrete model with SO(6)/SO(5) coset developed in \cite{gripaios} and clarify our specifications for the model.

In Section \ref{sec:ewclonld} we extract the scalar divergences of the theory with our previously developed techniques \cite{lindnermuzakka2022}.
Assuming one of the two scalars to be heavy enough, we integrate it out in order to match the resulting Lagrangian to the electroweak chiral Lagrangian with one light Higgs.

We conclude with a discussion of opportunities the model at hand offers in the outlook in Section \ref{sec:outlook}.

\section{The SO(6)/SO(5) Composite Higgs Model}
\label{unitarization}
\subsection{Unitarization}

The symmetry breaking scale of the new strong dynamics $ f $ in combination with the electroweak scale $v$ determines the most important parameter of CHMs:

\begin{equation}
\xi = \frac{v^2}{f^2} \ .
\end{equation}
Not only does it parameterize departures from SM interactions;
since it quantifies the splitting of the scales of spontaneous symmetry breaking, it is also crucial for setting up an effective theory at the electroweak scale.

In CHMs the Higgs sector is governed by strong interactions.
At the scale $ f $ we therefore have an effective description in terms of loops, or, equivalently, chiral dimensions \cite{power}.
To obtain the physics at the electroweak scale $ v $, the physics at $ f $ has to be integrated out.

We identify the following regimes.
\begin{itemize}
    \item In the limit $ \xi \rightarrow 0 $, the new physics entirely decouples and the SM is recovered, with a Standard Model effective field theory.

\end{itemize}

Otherwise, depending on the value of the ratio of the scales, there are two scenarios.

\begin{itemize}
    \item If $ \xi \ll 1 $ (at least small enough for an expansion to be meaningful) the physics at $ v $ is described by a \textit{double expansion} in canonical dimensions in powers of $ \xi $ and a loop expansion \cite{silh}.
    
    \item If $ f $ is not much higher than $ v $, the physics at $ f $ cannot be integrated out.
This scenario can actually be excluded with experimental data; we have $ \xi \lesssim 0.1 $.
The non-observation of new physics so far indicates a sufficiently large mass gap.
So in general we can have, in the effective approach, both an expansion in loops and canonical dimensions.
\end{itemize}

In the Standard Model the most important purpose of the linearly transforming Higgs doublet is that it fully unitarizes gauge boson scattering.
At high energies, the scattering of longitudinally polarized gauge bosons corresponds to the sole scattering of the eaten Goldstones, according to the \textit{Goldstone Boson Equivalence Theorem} \cite{equivalencetheorem1,equivalencetheorem2,equivalencetheorem3}.
The electroweak chiral Lagrangian without a Higgs scalar contains derivative interactions of the Goldstones (longitudinal components of the vector bosons) that violate perturbative unitarity at the cut-off scale $ \Lambda = 4\pi v $.
New physics would have to set in just there to unitarize the theory.

The electroweak chiral Lagrangian with a light Higgs scalar resonance, where the Higgs is not necessarily the SM one, is nonrenormalizable in the classical sense.
The Higgs resonance can at most partly unitarize the theory.
In the minimal composite Higgs model, the violation of perturbative unitarity is postponed to the scale $ \Lambda = 4\pi v/\sqrt{\xi} $ \cite{contino}, which is typical for any composite Higgs model \cite{dm}.
The other resonances of the strong sector take care of the full unitarization.

Summing up, in the case of $ \xi \rightarrow 0 $, new states are infinitely heavy and we stay with the SM, where the Higgs alone unitarizes the theory.
For $ \xi \rightarrow 1 $, which is excluded, new states have to ensure unitarity.
In composite Higgs scenarios they correspond to the resonances of the strong sector.
In between these extrema, unitarization is an interplay of Higgs and resonance mediations.

\subsection{Relation to Electroweak Chiral Theories}
Since the gauge boson couplings are measured to a high precision, the measurement of the Higgs couplings will hopefully serve as portal to new physics.
The Higgs couplings are within a few percent as in the Standard Model \cite{higgslhc,Chang2021} and
Higgs effective field theory should serve as the framework capturing any deviation from the SM as an indirect hint to new physics.
Any Lagrangian can in principle be matched to the effective framework at the electroweak scale $ v $ \cite{addsinglet, silh}.
The matching makes the modified phenomenology in the Lagrangian, compared to the Standard Model, directly visible \cite{higgscoupling, fithiggs}.
The electroweak chiral Lagrangian with a light Higgs can accommodate for various scenarios of electroweak symmetry breaking because of its general nature.
It does not need a UV-completion for itself to be of use but serves as a broad effective framework that can be used to particularize to specific models \cite{chiralnlo}.
As in CHMs the observed scalar can arise as a pseudo Goldstone from a larger symmetry breaking or just be an additionally included scalar.
In fact, an infinite number of coefficients in the polynomial functions of the Higgs field reflects a composite nature of the Higgs, where the internal structure cannot be described by a finite number of terms \cite{chiralnlo}.
The non-linear interaction structure is in clear contrast to the SM.

\subsection{The next-to-minimal Composite Higgs Model}
\label{NMCHM}

The coset $ SO(6)/SO(5) $ contains five Goldstone bosons.
Three of them get eaten by the $W$'s and the $Z$ and the other two form the observed Higgs and the additional scalar.
The Goldstones in the coset $ SO(6)/SO(5) $ transform in the fundamental \textbf{5} representation of $ SO(5) $\footnote{This is a general statement. The Goldstones in a coset $ SO(N)/SO(N-1) $ transform as fundamental SO(N-1) \cite{Wulzer}.}, which decomposes as $ \mathbf{4} \oplus \mathbf{1} $ under the subgroup $ SO(4) $, or as $ (\mathbf{2},\mathbf{2}) \oplus (\mathbf{1},\mathbf{1}) $ under $ SU(2)_L \times SU(2)_R $, see \ref{SO4SU2}.
The electroweak Higgs bi-doublet is accordingly associated with the bi-doublet $ (\mathbf{2},\mathbf{2}) $.
Hence, the theory contains an additional pseudo-scalar\footnote{The generators of $ SO(6) $ give rise to an automorphism acting as a CP transformation with $ h \rightarrow h $ and $ \eta \rightarrow - \eta $. There is no complex conjugation, since $ SO(6) $ is a real group \cite{niehoff}. For further reading, see \cite{Trautner,Grimus}} singlet $ \eta $.
We deduce the pseudo-scalar nature from its effective couplings to fermions below.

In Section \ref{fermion} it is shown that we have to enlarge the strong sector by a $ U(1) $ symmetry, which is gauged, to reproduce the correct hypercharge for fermions, as it is done in the minimal composite Higgs model \cite{contino}.
As this symmetry is not spontaneously broken, it does not affect the coset structure that determines the main features of the model.

Since the SM gauge group is identified as $ SU(2)_L \times U(1)_Y \subset SU(2)_L \times SU(2)_R $, with $ U(1)_Y $ generated by $ T_R^3 $, see \ref{SO4SU2} (neglecting the $ U(1)_X $ symmetry for now -- the Goldstones are uncharged under $ U(1)_X $ anyways), the singlet is a \textit{gauge singlet}.
It follows that also its shift-symmetry is not broken by SM gauging.
We are going to see that it can only obtain a potential via a suitable embedding of the right-handed fermions.

\subsection{The Goldstone Kinetic Lagrangian}
\label{so6goldstone}
We construct the kinetic term of the Lagrangian for the Goldstones along the way of the CCWZ construction \cite{coleman,callan}.
The Goldstone field is given by

\begin{equation}
\mathcal{U} = \exp\left(\frac{i\sqrt{2}\hat{h}^{\hat{a}}T^{\hat{a}}}{f}\right) \ .
\label{eqn:U}
\end{equation}
It contains scalar fields $\hat{h}$ and the coset generators of $T^{\hat{a}}$ of $SO(6)/SO(5)$ which are given in \ref{SO6}. 
The Lagrangian is given by

\begin{equation}
\mathcal{L}_{\text{kin}} = \frac{f^2}{2} \left(D_\mu \Sigma\right)^\dagger D^\mu \Sigma \ ,
\end{equation}
with

\begin{equation}
\Sigma  =\exp\left(\frac{i\sqrt{2}\hat{h}^{\hat{a}}T^{\hat{a}}}{f}\right) (0,0,0,0,0,1)^T =\frac{\sin\left(h'/f\right)}{h'}\left(\hat{h}_1,\hat{h}_2,\hat{h}_3,\hat{h}_4,\hat{h}_5,h'\cot \left(h'/f\right)\right)^T \ ,
\label{eqn:Sigma}
\end{equation}
where it is defined

\begin{equation}
\begin{aligned}
& h'^2 = \hat{h}_1^2+\hat{h}_2^2+\hat{h}_3^2+\hat{h}_4^2+\hat{h}_5^2\doteq \hat{h}^2+\eta^2 ,\\
& h' = \sqrt{h'^2} \ ,\\
& \hat{h}^2=\hat{h}_1^2+\hat{h}_2^2+\hat{h}_3^2+\hat{h}_4^2,\,\, \eta=\hat{h}_5.
\end{aligned}
\label{eqn:originalGoldstones}
\end{equation}
The electroweak covariant derivative acts on $\Sigma$ as

\begin{equation}
D_{\mu} \Sigma = \partial_{\mu} \Sigma + ig W_{\mu}^a \frac{\sigma^{a}}{2} \Sigma - ig' \Sigma \frac{\sigma^{3}}{2} B_{\mu} \ .
\label{eqn:covderiv}
\end{equation}
By going to unitary gauge in (\ref{eqn:Sigma}), setting $ \hat{h}_1=\hat{h}_2=\hat{h}_3=0 $ -- they are eaten by the gauge bosons -- and performing the field redefinitions

\begin{equation}
\hat{h}_4 = h' \ \cos\frac{\psi}{f} \ , \quad \eta = h' \sin \frac{\psi}{f} \ ,
\end{equation}
one arrives at

\begin{equation}
\Sigma = (0,0,0, \sin \frac{h'}{f} \cos \frac{\psi}{f}, \sin \frac{h'}{f} \sin \frac{\psi}{f} , \cos \frac{h'}{f})^T \ .
\end{equation}
The definitions

\begin{equation}
h= f \sin \frac{h'}{f} \cos \frac{\psi}{f} \ , \quad s= f \sin \frac{h'}{f} \sin \frac{\psi}{f} \ ,
\label{eqn:positive}
\end{equation}
are defined such that we have

\begin{equation}
\Sigma = \left(0,0,0, \frac{h}{f}, \frac{s}{f}, \frac{1}{f}\sqrt{f^2-h^2-s^2} \ \right)^T \ .
\label{eqn:Sigmanew}
\end{equation}
The kinetic term becomes

\begin{equation}
\begin{aligned}
\mathcal{L}_{\text{kin}} & = \frac{f^2}{2} \ |D_\mu \Sigma |^2
= \frac{1}{2} (\partial h)^2 +  \frac{1}{2} (\partial s)^2  + \frac{1}{2} \dfrac{(h\partial h+ s \partial s)^2}{f^2 - h^2- s^2} + \frac{g^2}{4} h^2 \left[ W^+W^- + \frac{1}{2 \cos^2 \theta_W} Z^2 \right] \ .
\label{eqn:goldkinetic}
\end{aligned}
\end{equation}
The electroweak mixing angle is $\cos \theta_W = e/g'$.
This is the result obtained in \cite{gripaios}.
Note that if we had chosen, e.g., $ h' $ and $ \psi $ as our fields, their kinetic term would have the geometrical structure of a 2-sphere:

\begin{equation}
\mathcal{L}_{\text{kin}} = (\partial h')^2 + \sin^2 \frac{h'}{f} (\partial \psi)^2 \ .
\label{eqn:sphere}
\end{equation}
Hence, there is no way to smoothly transform it into a plane, or in other words, to arrive at two canonical kinetic terms without derivative interactions of the two scalars as it was proposed in \cite{Redi}.
A direct matching is only possible to the most general form of the electroweak chiral Lagrangian with two scalars \cite{lindnermuzakka2022}.

It becomes clear from (\ref{eqn:goldkinetic}) that the electroweak scale is set by the vacuum expectation value (vev) of the field $ h $.
With the fact that $ h $ must obtain a vev and from the redefinition \eqref{eqn:positive}, the dangerously looking denominator in the third term on the right is constrained to be positive.
The other scalar $ s $ does not couple to the gauge fields.
There is thus no potential created for it by gauge boson loops.
In other words, it remains massless after the gauging, reflecting its nature as a \textit{gauge singlet}.

In the next section an explicit calculation of the effective Lagrangian for the SM gauge fields on the background of $ \Sigma $ reproduces the coupling in (\ref{eqn:goldkinetic}) with no coupling of $ s $ to the gauge fields.
This is because there is an unbroken shift symmetry among the original Goldstones in (\ref{eqn:Sigma}), namely the one associated to $ h_5 = \eta $.\footnote{The generator of this symmetry commutes with the SM generators, see the discussion in \ref{SO6}. There is however also the possibility of breaking the shift-symmetry by anomalies associated to the gauge bosons of the SM \textit{and} the strong sector \cite{chala,Sala}. We do not want to dive further into this aspect.}
This indicates that there has to be a physical, massless Goldstone left in the theory.
A potential and therefore a mass can be created for $ s $ by fermion loops, however, if the fermion embedding breaks $ SO(6) $ in such a way that also the $ U(1)_\eta $ shift symmetry is broken explicitly.
For this to happen, the right-handed fermions must not have a well-defined charge under $ U(1)_\eta $.
We discuss this in Section \ref{fermion}.

\subsection{The Gauge Boson Effective Lagrangian}	
\label{so6gauge}

The general $ SO(6) \times U(1)_X $ invariant effective Lagrangian (strong dynamics integrated out) for external gauge fields -- for invariance extended to a full gauging of $ SO(6) \times U(1)_X $; later only the SM gauge fields are kept -- on the background of $ \Sigma $ reads, in momentum space,

\begin{equation}
\mathcal{L}_{\text{g}} = \frac{1}{2} \mathcal{P}_T^{\mu\nu} \ \left[ \Pi_0^X (p^2) X_\mu X_\nu + \Pi_0(p^2) \langle A_\mu A_\nu \rangle + \Pi_1(p^2) \Sigma^T A_\mu A_\nu \Sigma \right] \ .
\label{eqn:gaugel}
\end{equation}
The gauge fields are denoted $ A_\mu = A_\mu^a T^a $, with the generators $ T^a $ spanning the whole group $ SO(6) $ and $ X_\mu $ the $ U(1)_X $ gauge field.
The form factors $ \Pi(p^2) $ contain the unknown integrated out dynamics of the strong sector.
The SM gauge group is the subgroup discussed above, generated by the corresponding generators given in (\ref{eqn:leftright}).
The hypercharge is defined as $ Y \doteq T_R^3 + Q_X $.
We need this definition to reproduce the correct hypercharge of the fermions.
Thus, the SM hypercharge gauge field is

\begin{equation}
\begin{aligned}
B_\mu & = \frac{g'}{g} W_{R\mu}^3 + \frac{g'}{g_X} X_\mu = \frac{g'}{g} W_{R\mu}^3 + \sqrt{1- \frac{g'^2}{g^2}} X_\mu  = \frac{g_X}{\sqrt{g^2+g_X^2}}W_{R\mu}^3 + \frac{g}{\sqrt{g^2+g_X^2}} X_\mu  \doteq s_X W_{R\mu}^3 + c_X X_\mu \, .
\end{aligned}
\end{equation}
We then find as the physical components of $ W_{R\mu} $ and $ X_\mu $,

\begin{equation}
W_{R\mu} |_{\text{phys}} = s_X B_\mu \, \quad X_\mu |_{\text{phys}} = c_X B_\mu \ .
\end{equation}
We use $ \Sigma $ as in (\ref{eqn:Sigmanew}) and find in (\ref{eqn:gaugel}), by keeping only the relevant gauge fields, the terms

\begin{equation}
\begin{aligned}
\langle A_\mu A_\nu \rangle & = \langle \left( W_{L\mu}^a T_L^a + W_{R\mu}^3 T_R^3 \right) \left( W_{L\nu}^a T_L^a + W_{R\nu}^3 T_R^3 \right) \rangle = \frac{1}{2} \left( W_{L\mu}^a W_{L\nu}^a + W_{R\mu}^3W_{R\nu}^3 \right) \, , \\
\Sigma^T A_\mu A_\nu \Sigma & =\Sigma^T   \left( W_{L\mu}^a T_L^a + W_{R\mu}^3 T_R^3 \right) \left( W_{L\nu}^a T_L^a + W_{R\nu}^3 T_R^3 \right) \Sigma 
\\
& = \frac{1}{4} \frac{h^2}{f^2} W_{L\mu}^a W_{L\nu}^a + \frac{1}{4} \frac{h^2}{f^2} W_{R\mu}^3 W_{R\nu}^3 - \frac{1}{2} \frac{h^2}{f^2} W_{L\mu}^3 W_{R\nu}^3 + ( \mu \leftrightarrow \nu ) \ .
\end{aligned}
\end{equation}
Keeping only the physical SM gauge fields, we obtain

\begin{equation}
\begin{aligned}
\mathcal{L}_{\text{gSM}}  = \frac{1}{2} P_T^{\mu\nu} \Big[ &\Pi_0^X(p^2) c_X^2 B_\mu B_\nu  + \frac{1}{2} \Pi_0(p^2) \left( W_{L\mu}^a  W_{L\nu}^a  + s_X^2 B_\mu B_\nu \right)\\
& + \frac{1}{2} \Pi_1(p^2) \frac{h^2}{f^2} \left(W_{L\mu}^a  W_{L\nu}^a + s_X^2 B_\mu B_\nu - 2 W_{L\mu}^3 s_X B_\nu  \right)    \Big] \, ,
\end{aligned} 
\end{equation}
or

\begin{equation}
\begin{aligned}
\mathcal{L}_{\text{gSM}} = \frac{1}{2} \mathcal{P}_T^{\mu\nu} \ & \left[ \left(  \Pi_0^X(p^2) c_X^2 + \frac{1}{2} \Pi_0(p^2) s_X^2 + \frac{1}{4} \Pi_1(p^2) \frac{h^2}{f^2} \ s_X^2 \right) B_\mu B_\nu \right. \\
& \quad + \left( \frac{1}{2} \Pi_0(p^2) + \frac{1}{4} \Pi_1(p^2)\frac{h^2}{f^2} \right) W_{L\mu}^a W_{L\nu}^a \\
& \left. \quad - \frac{1}{2} \Pi_1(p^2) \frac{h^2}{f^2} \ W_{L\mu}^3 \ s_X B_\nu \right] \ . 
\end{aligned}
\end{equation}
This is in accord with (\ref{eqn:goldkinetic}), since for zero momentum we have $ \Pi_1(0) = f^2 $, see the argument about the form factors $ \Pi(p^2) $ in \cite{contino}.

\subsection{The Fermion Embedding into SO(6)}
\label{fermion}

Considering the minimal possible fermion embeddings, the \textbf{4} representation of $ SO(6) $ is excluded because it does not contain a custodially symmetric (\textbf{2},\textbf{2}) subgroup, but decomposes as $ (\textbf{2},\textbf{1}) \oplus (\textbf{1},\textbf{2}) $ under $ SU(2)_L \times SU(2)_R $.
The \textbf{10} representation gives the fermions a well defined $ U(1)_\eta $ charge and thus there is no breaking of $ U(1)_\eta $, neither by gauge nor Yukawa interactions \cite{gripaios}.

We thus decide to put the fermions into the fundamental \textbf{6} representation, which decomposes as $ \mathbf{4} \oplus \mathbf{1} \oplus \mathbf{1} $ under $ SO(4) $, such that we can have the left-handed electroweak doublets in the custodial $ \mathbf{4}_c = (\mathbf{2},\mathbf{2})_c $\footnote{See \ref{SO4SU2} for details.}, as demanded for a protection of $ Z\rightarrow b_Lb_L $ couplings from new physics corrections \cite{zbb,niehoff,Wulzer}.
The right-handed fermions go into linear combinations of the singlets.
With this embedding, the representations can be matched to the electroweak ones at low energies \cite{niehoff} as we demonstrate below.

We now discuss the interaction of the SM fermions with the strong sector.
Linear mixings with strong "partners" arise from \textit{partial compositeness} \cite{KaplanFermion,Wulzer}.
These interactions originate at the cut-off scale of the theory, $ \Lambda $, where all the fields are in the \textbf{6} representations.
In the following we restrict ourselves to third generation quarks, since they have the largest mixing with the strong sector.
Naming the strong partners of the left- and right-handed fields $ \tilde{\Psi}_L $ and $ \Psi_R $, we get an interaction at the resonance scale of the form

\begin{equation}
\mathcal{L}_{\text{int},\Lambda} = \lambda_{tL} \bar{Q}_{tL} \Psi_{tR} + \lambda_{bL} \bar{Q}_{bL} \Psi_{bR} + \lambda_{tR} \bar{T}_R \tilde{\Psi}_{tL} + \lambda_{bR} \bar{B}_R \tilde{\Psi}_{bL} + \text{h.c.}\ .
\end{equation}
The fields $ Q_L, T_R, B_R $ denote the doublet and singlet fields of the SM and formally transform in the fundamental $SO(6)$ representation, as do their partners, such that the interaction Lagrangian is $G$-invariant.

The global symmetry breaking at lower energies is achieved by embedding the SM quarks into incomplete representations of $ SO(6) $ in the way explained above.
From their embedding properties we deduce, in terms of the fundamental SM fields $ b_L,t_L,b_R,t_R $:

\begin{equation}
\begin{aligned}
& T_R = \frac{1}{\sqrt{1+\epsilon_t^2}} \ (0,0,0,0,i\epsilon_t t_R, t_R)^T \, ,\\
& B_R = \frac{1}{\sqrt{1+\epsilon_b^2}} \ (0,0,0,0,i\epsilon_b b_R, b_R)^T \, ,\\
& Q_{tL} = \frac{1}{\sqrt{2}} \ (-i b_L, -b_L, -it_L, t_L,0,0)^T \, , \\
& Q_{bL} = \frac{1}{\sqrt{2}} \ (-i t_L, t_L, ib_L, b_L,0,0)^T \, .
\end{aligned}
\end{equation}
The embeddings of right-handed quarks contain a complex phase.
The parameter $ \epsilon $ can be chosen to be real, see \cite{Redi}, andobeys the restriction $ -1 \leq \epsilon \leq 1 $.
The doublet embeddings stem from (\ref{eqn:4cvector}).
Note that for $ \epsilon = \pm 1 $ the embedding of the right-handed quarks is $ SO(2)_\eta $ symmetric.
This symmetry is in that case not broken and right-handed fermions can be assigned a concrete charge under $ U(1)_\eta $.
As a consequence there is no fermion-loop induced potential for $ s $, see also \ref{SO6}.
The other extreme  $ \epsilon = 0 $ is also of phenomenological interest.
We come back to this case below.

For the definitions of $ Q_{tL} $ and $ Q_{bL} $, we projected out the fields with $ T_R^3 $ charge $ \mp 1/2 $, compare to (\ref{eqn:4cvector}).
This leads to the assignment of the additional $ U(1)_X $ symmetry charge for fermions:
With the hypercharge $ Y = T_R^3 + Q_X $, it follows that $ Q_X $ must be $ +2/3 $ for up-type quarks and $ -1/3 $ for down-type quarks \cite{niehoff}.
This way we find the correct SM hypercharge for the doublet $ (Y=1/6) $ and the singlets ($ Y=2/3, -1/3 $).
This embedding can thus be matched to the SM at the electroweak scale.

We rewrite the interaction Lagrangian in terms of the explicit SM fields, the left-handed doublet $ q_L = (t_L,b_L)^T $ and the right-handed fields $ t_R,b_R $.
For this, we introduce spurion fields $\Delta$ that make the connection.
We get

\begin{equation}
\begin{aligned}
\mathcal{L}_{\text{int},\Lambda} & =  \lambda_{tL} \bar{q}_L \Delta_{tL} \Psi_{tR} + \lambda_{bL} \bar{q}_L \Delta_{bL} \Psi_{bR}  + \lambda_{tR} \bar{t}_R \Delta_{tR} \tilde{\Psi}_{tL} + \lambda_{bR} \bar{b}_R \Delta_{bR} \tilde{\Psi}_{bL} + \text{h.c.} 
\end{aligned}
\label{eqn:int}
\end{equation}
With a proper assignment of transformation behaviour under the SM group and the global $SO(6)$ for the spurion fields, the interaction (\ref{eqn:int}) is again invariant under $ G = SO(6) $ \cite{carmona}.
We deduce the following quantum numbers under $ \underbrace{SU(2)_L \times U(1)_Y}_{\text{SM}} \ \times \ \underbrace{SO(6) \times U(1)_X}_{\text{comp. sector}} $:

\begin{equation}
\begin{aligned}
& q_L \in (2_{1/6},1_0) \, , \ t_R \in (1_{2/3},1_0) \, , \ b_R \in (1_{-1/3},1_0) \ , \\
& \Psi_{tR} \in (1_0,6_{2/3}) \, , \ \Psi_{bR} \in (1_0,6_{-1/3}) \, , \\
& \tilde{\Psi}_{tL} \in (1_0,6_{2/3}) \, , \ \tilde{\Psi}_{bL} \in (1_0,6_{-1/3}) \, , \\
& \Delta_{tL} \in (2_{1/6}, \bar{6}_{-2/3}) \, , \ \Delta_{bL} \in (2_{1/6},\bar{6}_{1/3}) \, , \\ 
& \Delta_{tR} \in (1_{2/3}, \bar{6}_{-2/3}) \, , \ \Delta_{bR} \in (1_{-1/3},\bar{6}_{1/3}) \ .
\end{aligned}
\end{equation}
The physical (expectation) values of the spurion fields are given by

\begin{equation}
\begin{aligned}
& \langle \Delta_{t,bR} \rangle = \frac{1}{\sqrt{1+ \epsilon_{t,b}^2}} (
0, 0, 0, 0,i  \epsilon_{t,b}, 1)
\, ,
\\
& \langle \Delta_{tL} \rangle = \frac{1}{\sqrt{2}} \begin{pmatrix}
0 & 0 & -i & 1 & 0 & 0 \\
-i & -1 & 0 & 0 & 0 & 0
\end{pmatrix} \, ,
\\
& \langle \Delta_{bL} \rangle = \frac{1}{\sqrt{2}} \begin{pmatrix}
-i & 1 & 0 & 0 & 0 & 0 \\
0 & 0 & i & 1 & 0 & 0
\end{pmatrix} \, .
\end{aligned}
\end{equation}
These "vevs" break the global symmetry and consequently the mixings create a potential for the Goldstone Higgs.
The form of the potential can thus be obtained via a spurion analysis \cite{carmona}.
We construct the invariants with the spurion fields and the Goldstone field $ \mathcal{U} $ (\ref{eqn:U}).
Setting the spurions to their vevs yields a non-trivial structure of the potential.
Note that the strength of the breaking of $ G $ is determined by the mixing parameters $ \lambda $ in (\ref{eqn:int}).
The mixings to the strong sector are responsible for the generation of masses for the fermions and from considerations on partial compositeness, we can deduce that the breaking is weak, except maybe for the top quark, and use these parameters as expansion parameters \cite{KaplanFermion,Wulzer}.

The idea of building the invariant structures stems from the CCWZ construction \cite{coleman,callan}.
The Goldstone field $ \mathcal{U} $ is a local element of $ G $ with transformation property $ \mathcal{U} \rightarrow \mathbf{g} \, \mathcal{U} \, \mathbf{h}^{-1} $, where $ \mathbf{g} \in SO(6) \ , \ \mathbf{h} \in SO(5) $.
$ \mathcal{U}^{-1} $ can thus be used to act on fields with $ SO(6) $ indices, like the spurion field $ \Delta $, to get a "dressed" one transforming with $ \mathbf{h} $.
The dressed object can be split into several $ SO(5) $ irreducible representations.
Since, however, the spurion fields simultaneously transform under the SM gauge group, we have to keep the invariance in that respect, too.
The spurion fields can therefore only enter in the combinations $ \Delta_L^\dagger \Delta_L $ and $ \Delta_R^\dagger \Delta_R $.
\\
We identify as invariants the following singlet-singlet (66-) components,

\begin{equation}
\begin{aligned}
\mathcal{O}(\lambda^2): & \quad (\mathcal{U}^\dagger \Delta_{tL}^\dagger \Delta_{tL} \mathcal{U})_{66} = \Sigma^t \Delta_{tL}^\dagger \Delta_{tL} \Sigma =  \frac{1}{2} \frac{h^2}{f^2} \ , \\
& \quad (\mathcal{U}^\dagger \Delta_{tR}^\dagger \Delta_{tR} \mathcal{U})_{66}= \frac{1}{1+\epsilon^2}\frac{|\phi|^2}{f^2} \ , \\
\mathcal{O}(\lambda^4): & \quad (\mathcal{U}^\dagger \Delta_{tL}^\dagger \Delta_{tL} \mathcal{U})^2_{66} = \frac{1}{4} \frac{h^4}{f^4} \ , \\
& \quad (\mathcal{U}^\dagger \Delta_{tR}^\dagger \Delta_{tR} \mathcal{U})^2_{66}=\frac{1}{(1+\epsilon^2)^2}\frac{|\phi|^4}{f^4} \ , \\
& \quad (\mathcal{U}^\dagger \Delta_{tL}^\dagger \Delta_{tL} \mathcal{U})_{66} \ (\mathcal{U}^\dagger \Delta_{tR}^\dagger \Delta_{tR} \mathcal{U})_{66} = \frac{1}{2} \frac{1}{1+\epsilon^2} \frac{h^2}{f^2} \frac{|\phi|^2}{f^2} \ ,
\end{aligned}
\label{eqn:SO6invariants}
\end{equation}
where $ \phi= \sqrt{f^2-s^2-h^2} +i \epsilon_t s $.
For the b quark, the structures are the same, but the impact of the top is overwhelming.
Also the gauge contributions to the potential are overwhelmed by the large Yukawa coupling of the top \cite{Wulzer}.
Collecting the terms, we find for the form of the potential:

\begin{equation}
V(h,s) = \alpha h^2 + \kappa h^4 + |\phi|^2 ( \beta + \gamma h^2 + \delta |\phi|^2) \ ,
\label{eqn:SO6potential}
\end{equation}
in accordance with the form found via a calculation of the Coleman-Weinberg potential in \cite{gripaios}.
Note that indeed there is no potential for $ s $ if $ \epsilon = \pm 1 $.
\\
In the framework of partial compositeness the interactions of fermions with Goldstones is mediated by mixing with the fermionic partners in the strong sector.
Integrating them out formally we obtain the low energy Yukawa couplings through the form factors by coming up with the most general $ SO(6) $ invariant Lagrangian of fermion-Goldstone interaction:

\begin{equation}
\begin{aligned}
\mathcal{L}_{\text{f}} & = \bar{Q}_{tL}^i \, \slashed{p} \left( \delta^{ij} \Pi_0^{tL} (p^2)  + \Sigma^i \Sigma^j \Pi_1^{tL} (p^2) \right) Q_{tL}^j \\
& + \bar{Q}_{bL}^i \, \slashed{p} \left( \delta^{ij} \Pi_0^{bL} (p^2)  + \Sigma^i \Sigma^j \Pi_1^{bL} (p^2) \right) Q_{bL}^j \\
& + \bar{T}_R^i \slashed{p} \left( \delta^{ij} \Pi_0^{tR} (p^2)  + \Sigma^i \Sigma^j \Pi_1^{tR} (p^2) \right) T_R^j \\
& + \bar{B}_R^i \slashed{p} \left( \delta^{ij} \Pi_0^{bR} (p^2)  + \Sigma^i \Sigma^j \Pi_1^{bR} (p^2) \right) B_R^j \\
& + \bar{Q}_{tL}^i \left( \delta^{ij} \hat{M}_0^{t} (p^2)  + \Sigma^i \Sigma^j \hat{M}_1^{t} (p^2) \right) T_R^j \\
& + \bar{Q}_{bL}^i \left( \delta^{ij} \hat{M}_0^{b} (p^2)  + \Sigma^i \Sigma^j \hat{M}_1^{b} (p^2) \right) B_R^j \ + \text{h.c.}
\end{aligned}
\end{equation}
Since we are mainly interested in the effective Goldstone-fermion interactions we extract their structure similarly to \eqref{eqn:SO6invariants} and get:

\begin{equation}
\begin{aligned}
\mathcal{L}_{\text{f,G}} & = \Pi_1^{tL} (p^2)  \, \bar{t}_L \slashed{p} \, t_L \, \frac{h^2	}{2f^2} +  \Pi_1^{bL} (p^2)  \, \bar{b}_L \slashed{p} \, b_L \, \frac{h^2	}{2f^2} \\
& +  \Pi_1^{tR} (p^2)  \, \bar{t}_R \slashed{p} \, t_R \, \frac{1}{1+\epsilon^2} \frac{ |\phi|^2	}{f^2} +\Pi_1^{bR} (p^2)  \, \bar{b}_R \slashed{p} \, b_R \, \frac{1}{1+\epsilon^2} \frac{ |\phi|^2	}{f^2} \\
& + \hat{M}_1^t (p^2) \bar{t}_L t_R \, \frac{1}{\sqrt{2+2\epsilon^2}}\frac{h \phi}{f} + \hat{M}_1^b (p^2) \bar{b}_L b_R \, \frac{1}{\sqrt{2+2\epsilon^2}} \frac{h \phi}{f} \ +\text{h.c.}
\end{aligned}
\end{equation}
where, again, $ \phi= \sqrt{f^2-s^2-h^2} +i \epsilon_{t,b} s $.
Specifically, Yukawa couplings of the form

\begin{equation}
y_{hs} \sim \ \bar{t} \gamma^5 t \, \frac{h\, is}{f} 
\label{eqn:yuks}
\end{equation}
are obtained.
We can thus define $ s $ as a pseudo-scalar under parity.
Note that there are no such couplings if $ \epsilon = 0 $.\\

In \cite{chala} the next-to-minimal CHM with the fermions in the \textbf{20} representation of $ SO(6) $ is considered as special case of a general EFT for the minimally extended strongly interacting light Higgs.
In this representation, the potential can lead to EWSB already at leading order of the spurion expansion.

\section{One-Scalar-Loop Divergence via Matching to the Electroweak Chiral Lagrangian}\label{sec:ewclonld}
The electroweak chiral Lagrangian with a light Higgs (EWCLh) is to leading order given in compact form by \cite{completerenorm}

\begin{equation}
\begin{aligned}
\mathcal{L} _{ew\chi}= & -\frac{1}{4}G_{\mu\nu}^{i2}-\frac{1}{4}B_{\mu\nu}^2-\frac{1}{4}W_{\mu\nu}^{a2}+\frac{1}{2}\partial_\mu h\,\partial^\mu h-V(h) \\
& +\frac{v^2}{4}\langle (D_\mu U)^\dagger D^\mu U \rangle F(h)+\bar{\psi}i \slashed{D}\psi-\bar{\psi}m(h,U)\psi \ ,
\end{aligned}
\label{eqn:EWCLhcompact}
\end{equation}
where $G^i_{\mu\nu}$, $W_{\mu\nu}$, and $B_{\mu\nu}$ are the gauge field strength tensors of the strong and electroweak interactions and the covariant derivative is given by

\begin{equation}
D_\mu U=\partial_\mu U+igW_\mu U-ig'B_\mu UT^3
\end{equation}
and

\begin{equation}
U=\exp\left(2i\varphi/v\right) \, , \quad \textrm{with} \quad \varphi = \varphi^a \ T^a \ ,
\end{equation} is the electroweak Goldstone matrix.
The $ SU(2) $ generators $ T^a $ are normalized such that

\begin{equation}
\langle T^a T^b \rangle = \frac{1}{2} \delta^{ab} \  \Rightarrow T^a = \frac{\sigma^a}{2} \ ,
\end{equation} 
with the Pauli sigma matrices $\sigma^a$.
The Yukawa term $ m(h,U) $ is short for

\begin{equation}
m(h,U) \doteq UM(h) P_R + M^\dagger(h,s) U^\dagger P_L \ ,
\end{equation}
with the mass matrix $ M $ and the fermion field $ \psi $ defined as (generation indices suppressed)

\begin{equation}
M(h) = \text{diag} (M_u(h),M_d(h),M_\nu(h),M_e(h)) \ ,\quad \psi= (u,d,\nu,e)^T \ .
\label{eqn:MandPsi}
\end{equation}
Note that $U$ is a 2 x 2 matrix and $M(h)_{ij}^\alpha $ as defined in (\ref{eqn:MandPsi}) is a 4 x 4 matrix. 
Since we are working with SM particle content, we have $ M_\nu = 0 $.
\\		
Further, we have the expansions of the Higgs-dependent functions

\begin{equation}\label{eq:polynoms}
\begin{aligned}
F(h) = 1+ \sum_{n=1}^\infty F_n \left( \frac{h}{v} \right)^n  , \, V(h) = v^4 \sum_{n=2}^\infty V_n \left( \frac{h}{v} \right)^n  , \,  M(h) = \sum_{n=0}^\infty M_n \left( \frac{h}{v} \right)^n  .
\end{aligned}
\end{equation}

\subsection{Recap: The Background Field Method}

The background field method provides a convenient way of computing the one-loop effective action \cite{thooft,Abbott,Ramond,dewittquantum}.
The one-scalar-loop divergence of the electroweak chiral Lagrangian with a light Higgs was performed in \cite{guo} and the complete one-loop renormalization in \cite{completerenorm}.
Unitary gauge is chosen for the Goldstones such that their background contribution vanishes during the calculation \cite{Dittmaier1,Dittmaier2}.

In general, by separating the scalar field configurations into a classical background and quantum fluctuations around it,

\begin{equation}
\phi = \phi_c + \phi_q \ ,
\end{equation}
it is obtained:

\begin{equation}\label{eq:Seff}
S_{\text{eff}}^{\text{1loop}} = \frac{i}{2} \textrm{tr} \ \int d^4x \int \frac{d^4p}{(2\pi)^d} \ \ln \ \Delta(x,\partial_x+ip) \ ,
\end{equation}
where tr is the trace only over operator space and the differential operator $\Delta(x,\partial_x+ip)$ is given by

\begin{equation}
\dfrac{ \delta^2 \mathcal{L}}{\delta \phi(x) \delta \phi(y)} \bigg|_{\phi_c} \doteq  \Delta(x,\partial_x) \ \delta^4(x-y) \, .
\label{eqn:diffop}
\end{equation}
The differential operator can be expressed in the general form

\begin{equation}
\Delta(x,\partial_x) = - \mathcal{D}_\mu \mathcal{D}^\mu - Y \, , \quad \mathcal{D}_\mu = \partial_\mu + X_\mu \, .
\label{eqn:diffopx}
\end{equation}
In these terms it is found

\begin{equation}
\Delta(x,\partial_x+ip) = p^2 \left(1- \dfrac{2ip \mathcal{D} + \mathcal{D}^2 +Y}{p^2} \right) \, .
\end{equation}
The logarithm in \eqref{eq:Seff} can be expanded and the integrals evaluated using dimensional regularization \cite{dimreg1,dimreg2}.
The t'Hooft master formula for the divergent terms of the effective Lagrangian that t'Hooft himself found using symmetry arguments \cite{thooft} is then given by

\begin{equation}
\mathcal{L}_{\text{eff,div}}^{\text{1loop}} = \dfrac{1}{32\pi^2 \epsilon} \text{tr} \left[ \frac{1}{12} X_{\mu\nu}X^{\mu\nu} + \frac{1}{2} Y^2 \right] \, ,
\label{eqn:tHooft}
\end{equation}
with

\begin{equation}
X_{\mu\nu} = [\mathcal{D}_\mu,\mathcal{D}_\nu] \, .
\end{equation}

\subsection{Recap: Renormalization of Kinetic Terms with two Singlets via Matrix Field Redefinitions}
As explained in Section \ref{so6goldstone}, it is not possible to get rid of derivative interactions of the two scalars.
In order to extract the divergences caused by the Higgs sector of the model we perform a matching to the electroweak chiral Lagrangian with two scalars.
We showed in our previous work how to renormalize scalar kinetic terms with two and any number of fields \cite{lindnermuzakka2022}.
To this end we have to transform the kinetic part of the Lagrangian by means of field redefinitions.
The case of the $SO(6)/SO(5)$ CHM serves as an exemplary theory to make use of our method for two scalar singlet fields.

First, we set up the most general form of a kinetic term for two scalar fields:

\begin{equation}
\mathcal{L}_{\text{kin,2}}=\frac{1}{2}\, (\partial h)^2(1+F_{11}(h,s))+\frac{1}{2}\,(\partial s)^2(1+F_{22}(h,s))+\partial h\, \partial s \,F_{12}(h,s) \ .
\label{eqn:generalkinetictwoscalar}
\end{equation}
When attempting to renormalize this form via the background field method directly, it leads to kinetic mixing of quantum fluctuations.
This leads to a differential operator $ \Delta $ with $ \partial^2 $-operators in the off-diagonal elements in contrast to the desired form given in \eqref{eqn:diffopx}.
We need matrix field redefinitions to obtain a properly defined differential operator.
\if false
We can rewrite \eqref{eqn:generalkinetictwoscalar} as

\begin{equation}
\mathcal{L}_{\text{kin}}=\frac{1}{2} ( \partial h , \, \partial s) \begin{pmatrix}
1+ F_{11} & F_{12} \\
F_{12} & 1+ F_{22}
\end{pmatrix}
\begin{pmatrix}
\partial h \\ \partial s
\end{pmatrix} \ .
\end{equation}
\fi
Defining $ h_i = (h,s)^T $, it is compactly written as

\begin{equation}
\mathcal{L}_{\text{kin,2}}=\frac{1}{2}\left(\delta_{ij}+F_{ij}(h)\right) \partial_\mu h_i \,\partial^\mu h_j,
\end{equation} 
where $F_{ij}$ is a symmetric matrix.
We separate the fields into classical background plus quantum fluctuations $h_i \rightarrow h_i +\tilde{h}_i $.
We need the Lagrangian at second order in the quantum fluctuations:

\begin{equation}
\begin{aligned}
&\mathcal{L}_{\text{kin,2}}  \overset{\mathcal{O}(\tilde{h}^2)}{=} \frac{1}{2}\left(\delta_{ij}+F_{ij}\right)\partial_\mu \tilde{h}_i \,\partial^\mu \tilde{h}_j + F'_{ij,k}(\partial_\mu h_i)\, \tilde{h}_k \partial^\mu \tilde{h}_j + \frac{1}{4}F''_{ij,kl} (\partial_\mu h_i)(\partial^\mu h_j) \tilde{h}_k \tilde{h}_l \\
& \qquad  \quad= -\frac{1}{2} \tilde{h}_i \Big[\left(\delta_{ij}+F_{ij}\right)\partial^2+\left(F'_{ij,k}-2F'_{jk,i}\right)(\partial_\mu h_k)\partial^\mu -\frac{1}{2}F''_{kl,ij}(\partial_\mu h_k)(\partial^\mu h_l) \Big] \tilde{h}_j \ .
\end{aligned}
\end{equation}
For any symmetric matrix $S^\mu$ and vector $\eta$ we have:

\begin{equation}
\eta^T S^\mu \partial_\mu \eta= -\frac{1}{2}\eta^T (\partial_\mu S^\mu)\eta +\text{total der.}
\end{equation}
With this we find

\begin{equation}
\begin{aligned}
\mathcal{L}_{\text{kin,2}}  &\overset{\mathcal{O}(\tilde{h}^2)}{=}-\frac{1}{2} \tilde{h}_i\Big[\left(\delta_{ij}+F_{ij}\right)\partial^2+\left(F'_{ik,j}-F'_{jk,i}\right)(\partial_\mu h_k)\partial^\mu \\
& \qquad\ -\partial_\mu\left[(F'_{ij,k}-F'_{jk,i}-F'_{ik,j})\partial^\mu h_k\right] -\frac{1}{2}F''_{kl,ij}(\partial_\mu h_k)(\partial^\mu h_l) \Big]\tilde{h}_j\\
& \equiv -\frac{1}{2} \tilde{h}^T \left[A(h)\partial^2+2B^\mu(h)\partial_\mu+C \right] \tilde{h} \ ,
\end{aligned}
\end{equation}
where 

\begin{equation}
\begin{aligned}
A_{ij} (h)&=\delta_{ij}+F_{ij}(h)\,,\quad B^\mu_{ij}(h)=\frac{1}{2}\left(F'_{ik,j}-F'_{jk,i}\right)(\partial_\mu h_k)\,,\\
C_{ij}(h)& = -\partial_\mu\left[(F'_{ij,k}-F'_{jk,i}-F'_{ik,j})\partial^\mu h_k\right] -\frac{1}{2}F''_{kl,ij}(\partial_\mu h_k)(\partial^\mu h_l) \ .
\end{aligned}
\end{equation}
As is desired for the method, $A$ and $C$ are symmetric, while $B^\mu$ is antisymmetric.
We express the components of $ B $ and $ C $ in terms of the background scalars $ h $ and $ s $.
Derivatives with respect to $ h $ are denoted with a prime and those with respect to $ s $ with a dot.
The components read

\begin{equation}
\begin{aligned}
B^\mu_{12} & = - B^\mu(h)_{21}= \frac{1}{2} (\dot{F}_{11} - F_{12}')\, \partial^\mu h+ (\dot{F}_{12} - F_{22}') \, \partial^\mu s \ , \\
C_{11} & = \frac{1}{2} F_{11}'' (\partial h)^2 + F_{12}'' (\partial h)(\partial s) + (-\ddot{F}_{11} + 2 \dot{F}_{12}' - \frac{1}{2} F_{22}'') (\partial s)^2  + F_{11}' (\partial^2 h) + (-\dot{F}_{11} + 2F_{12}') (\partial^2 s) \ , \\
C_{22} & = \frac{1}{2} \ddot{F}_{22} (\partial s)^2 + \ddot{F}_{12} (\partial h)(\partial s) + (-F_{22}'' + 2 \dot{F}_{12}' - \frac{1}{2} \ddot{F}_{11}) (\partial h)^2 + \dot{F}_{22} (\partial^2 s) + (-F'_{22} + 2\dot{F}_{12}) (\partial^2 h) \ , \\
C_{12} & = C_{21} = \frac{1}{2} \dot{F}_{11}' (\partial h)^2 +\frac{1}{2} \dot{F}_{22}' (\partial s)^2 + (F_{22}''+ \ddot{F}_{11}-\dot{F}_{12}') (\partial h)(\partial s)  + F_{22}' (\partial^2 s) + \dot{F}_{11} (\partial^2 h) \ .
\end{aligned}
\end{equation}
We want to obtain a differential operator in the standard form of Equation (\ref{eqn:diffopx}).
We therefore have to perform a matrix field rescaling

\begin{equation}
\tilde{h} \rightarrow A^{-1/2} \, \tilde{h} \ .
\end{equation}
Hence, we have 

\begin{equation}
\begin{aligned}
&\partial_\mu \tilde{h} \longrightarrow (\partial_\mu A^{-1/2}) \tilde{h}+ A^{-1/2}\partial_\mu \tilde{h} \ , \\
& \partial^2 \tilde{h} \longrightarrow (\partial^2 A^{-1/2})\tilde{h} + 2(\partial_\mu A^{-1/2})\partial^\mu \tilde{h} + A^{-1/2} \partial^2\tilde{h} \ .
\end{aligned}
\end{equation}
With this, we obtain 

\begin{equation}
\begin{aligned}
\mathcal{L}_{\text{kin,2}} &\overset{\mathcal{O}(\tilde{h}^2)}{=} -\frac{1}{2}\tilde{h}^T \Big[\partial^2+2\left(A^{1/2} (\partial^\mu A^{-1/2})+A^{-1/2} B^\mu A^{-1/2}  \right)\partial_\mu\\
& \qquad \ + A^{-1/2} CA^{-1/2}+A^{1/2} (\partial^2 A^{-1/2})+2A^{-1/2} B^\mu (\partial_\mu A^{-1/2}) \Big]\tilde{h} \ .
\end{aligned}
\end{equation}
Next, we antisymmetrize the coefficient of $\partial_\mu$ using

\begin{center}
$ A^{1/2} (\partial^\mu A^{-1/2}) = - (\partial^\mu A^{1/2})A^{-1/2} $ 
\end{center}
and symmetrize the constant terms again using partial integration.
We obtain -- remember that $B_\mu$ is antisymmetric and $C$ is symmetric --

\begin{equation}
\begin{aligned}
\mathcal{L}_{\text{kin,2}} &\overset{\mathcal{O}(\tilde{h}^2)}{=} -\frac{1}{2}\tilde{h}^T \Big[ \ \partial^2 +2\Big(A^{-1/2} B^\mu A^{-1/2} +\frac{1}{2}A^{1/2} (\partial^\mu A^{-1/2})- \frac{1}{2}(\partial^\mu A^{-1/2}) A^{1/2}  \Big)\partial_\mu\\
&\qquad \qquad \qquad+ A^{-1/2} CA^{-1/2} +A^{-1/2} B^\mu (\partial_\mu A^{-1/2}) -   (\partial_\mu A^{-1/2}) B^\mu  A^{-1/2} \\
&\qquad \qquad \qquad -\frac{1}{2}  \left(\partial_\mu A^{-1/2} \,\partial^\mu A^{1/2}+  \partial_\mu A^{1/2} \,\partial^\mu A^{-1/2}\right)\Big]\tilde{h}\\
&  \doteq -\frac{1}{2}\tilde{h}^T \left[(\partial_\mu+X_\mu)^2+Y \right]\tilde{h} \ ,
\end{aligned}
\end{equation}
with 

\begin{equation}
\begin{aligned}
X^\mu & = A^{-1/2} B^\mu A^{-1/2} +\frac{1}{2}A^{1/2} (\partial^\mu A^{-1/2})- \frac{1}{2}(\partial^\mu A^{-1/2}) A^{1/2} \ , \\
Y &=A^{-1/2} CA^{-1/2} +A^{-1/2} B^\mu (\partial_\mu A^{-1/2}) -   (\partial_\mu A^{-1/2}) B^\mu  A^{-1/2} \\
&\quad -\frac{1}{2}  \left(\partial_\mu A^{-1/2} \,\partial^\mu A^{1/2}+  \partial_\mu A^{1/2} \,\partial^\mu A^{-1/2}\right)-X_\mu X^\mu \ .
\end{aligned}
\label{eqn:XandY}
\end{equation}
These are well-defined objects for the formula of the differential operator (\ref{eqn:diffopx}) from which one obtains the divergence using the t'Hooft master formula (\ref{eqn:tHooft}).

\subsection{Matching of the next-to-minimal CHM to the Electroweak Chiral Lagrangian with two Higgs Fields}
For further processing in this section, we rename the fields:

\begin{equation}
h \longrightarrow \xi_0 \quad s \longrightarrow \eta_0 \, .
\end{equation}
We write down the embedding of the model into the electroweak chiral Lagrangian with two scalars, including the potential (\ref{eqn:SO6potential}):

\begin{equation}
\begin{aligned}
\mathcal{L} &=-\frac{1}{4} W_{\mu\nu}^{a2}-\frac{1}{4} B_{\mu\nu}^2 +\frac{\xi^2_0}{4}\left\langle (D_\mu U)^\dagger D^\mu U\right\rangle \\
& +\frac{1}{2}(\partial \xi_0)^2\left(1+\frac{\xi_0^2}{f^2-\xi_0^2-\eta_0^2}\right)+\frac{1}{2}(\partial \eta_0)^2\left(1+\frac{\eta_0^2}{f^2-\xi_0^2-\eta_0^2}\right)\\
&+\partial \xi_0 \,\partial \eta_0 \frac{\xi_0\eta_0}{f^2-\xi_0^2-\eta_0^2}- \left[A_1 \xi_0^2+A_2 \eta_0^2+B\,\xi_0^2\eta_0^2+C_1 \xi_0^4+C_2 \eta_0^4  \right] \\
& +\bar{\psi} m(\xi_0,\eta_0,U) \psi
\label{eqn:electroweakSO6}
\end{aligned}
\end{equation}
The potential permits both scalars to obtain a vev, so for generality we write

\begin{equation}
\xi_0\rightarrow v +\tilde{\xi}_0 \ , \quad \eta_0\rightarrow v_\eta +\tilde{\eta}_0 \, .
\end{equation}
$v$ has to be identified with the electroweak scale.
This renders the kinetic terms unnormalized.
So we diagonalize the kinetic terms with a mixing- or canonicalization-matrix 

\begin{equation}
\begin{aligned}
\mathcal{C} &\doteq \begin{pmatrix}
1+\frac{v^2}{f^2-v^2-v_\eta^2} & \frac{v v_\eta}{f^2-v^2-v_\eta^2}\\
\frac{v v_\eta}{f^2-v^2-v_\eta^2} &1+\frac{v_\eta^2}{f^2-v^2-v_\eta^2}
\end{pmatrix} =\mathcal{R}(v_\eta, v) \,\, \text{diag}\left(\frac{f^2}{f^2-v^2-v_\eta^2}, 1 \right) \,\, \mathcal{R}^T(v_\eta, v)\\
&=\frac{1}{\sqrt{1+v^2_\eta/v^2}} \begin{pmatrix}
1 & \frac{v_\eta}{v} \\ \frac{v_\eta}{v} &-1
\end{pmatrix}\begin{pmatrix}
\frac{f^2}{f^2-v^2-v_\eta^2} & 0\\0 & 1
\end{pmatrix}\frac{1}{\sqrt{1+v^2_\eta/v^2}} \begin{pmatrix}
1 & \frac{v_\eta}{v} \\ \frac{v_\eta}{v} &-1
\end{pmatrix} \ .
\end{aligned}
\end{equation}
Note that the whole expression is well defined in the limit $v_\eta=0$.\footnote{The matrix $\mathcal{R}$ represents an improper rotation, which is equivalent to a rotation followed by a mirror transformation.}
To properly normalize the Lagrangian, we define 

\begin{equation}
\begin{aligned}
\begin{pmatrix}
\xi \\ \eta
\end{pmatrix} &=\frac{1}{\sqrt{1+v^2_\eta/v^2}}\begin{pmatrix}
\frac{f}{\sqrt{f^2-v^2-v_\eta^2}} & 0\\0 & 1
\end{pmatrix} \begin{pmatrix}
1 & \frac{v_\eta}{v} \\ \frac{v_\eta}{v} &-1
\end{pmatrix}\begin{pmatrix}
\xi_0 \\ \eta_0
\end{pmatrix}\\
& =\frac{1}{\sqrt{1+v^2_\eta/v^2}}\begin{pmatrix}
\frac{f}{\sqrt{f^2-v^2-v_\eta^2}} & \frac{fv_\eta}{v\sqrt{f^2-v^2-v_\eta^2}}\\\frac{v_\eta}{v} & -1
\end{pmatrix} \begin{pmatrix}
\xi_0 \\ \eta_0
\end{pmatrix}\doteq \mathcal{R} \begin{pmatrix}
\xi_0 \\ \eta_0
\end{pmatrix} \ .
\end{aligned}
\end{equation}
The Lagrangian (\ref{eqn:electroweakSO6}) can now be rewritten in terms of canonical fields by shifting them:

\begin{equation}
\begin{aligned}
\begin{pmatrix}
\xi_0 \\ \eta_0
\end{pmatrix}=\begin{pmatrix}
v\\v_\eta
\end{pmatrix}+\mathcal{R}^{-1} \begin{pmatrix}
\xi \\ \eta
\end{pmatrix} \, ,
\end{aligned}
\end{equation}
with

\begin{equation}
\mathcal{R}^{-1} = \frac{1}{\sqrt{1+v^2_\eta/v^2}} \begin{pmatrix}
\frac{\sqrt{f^2-v^2-v_\eta^2}}{f} & \frac{v_\eta}{v}\\ \frac{v_\eta}{v} \frac{\sqrt{f^2-v^2-v_\eta^2}}{f} & - 1
\end{pmatrix} \ .
\end{equation}
So that,

\begin{equation}
\begin{aligned}
\xi_0 &= v+ \frac{\sqrt{f^2-v^2-v_\eta^2}}{f\sqrt{1+v_\eta^2/v^2}} \, \xi + \frac{v_\eta}{v}\frac{1}{\sqrt{1+v_\eta^2/v^2}} \, \eta \ , \\
\eta_0 &= v_\eta + \frac{v_\eta}{v} \frac{\sqrt{f^2-v^2-v_\eta^2}}{f\sqrt{1+v_\eta^2/v^2}} \, \xi - \frac{1}{\sqrt{1+v_\eta^2/v^2}} \, \eta \ .
\end{aligned}
\end{equation}
In general, defining

\begin{equation}
\phi_{0i} = ( \xi_0 , \, \eta_0) \, ,
\end{equation}
we have for the scalar sector:

\begin{equation}\label{eqn:primi}
\begin{aligned}
\mathcal{L} & \supset \frac{\phi_{01}^2}{4}\langle L_\mu L^\mu\rangle +\frac{1}{2} \partial \phi_{0i} \,\partial \phi_{0j} \left(\delta_{ij}+f_{ij}(\phi_0)\right) - \left[A_{1} \phi_{01}^2 + A_{2} \phi_{02}^2+B \phi_1^2\phi_{02}^2+C_1 \phi_{01}^4+ C_2 \phi_{02}^4 \right] \,.
\end{aligned}
\end{equation}
Let us denote the vevs as

\begin{equation}
\left(\langle \phi_{01}\rangle, \, \langle \phi_{02} \rangle \right)=\left(v,\, v_\eta \right) \doteq \left(v_1,\, v_2\right) \, .
\end{equation}
Since vevs are the global minima of the potential, we have 

\begin{equation}
\begin{aligned}
A_1 = -B v_2^2-2C_1 v_1^2 \ , \\
A_2 = -B v_1^2-2C_2 v_2^2 \ .
\end{aligned}\label{eqn:vevrelation}
\end{equation}
Note that to obtain the second line we had to assume $ v_2 \neq 0 $.
This is crucial for the computations ahead, see also the discussion at the end of Section \ref{integrateout}.
Expanding the fields around their vevs and plugging them into the original Lagrangian, we find that the kinetic terms need to be normalized:

\begin{equation}
\begin{aligned}
& \frac{1}{2} \partial \phi_{0i} \,\partial \phi_{0j} \left(\delta_{ij}+f_{ij}(\phi_0)\right)   \overset{\phi_{0i} = v_i+\tilde{\phi_{0i}}}{=} \frac{1}{2}\partial\tilde{\phi}_{0i}\partial\tilde{\phi}_{0j} \left[\delta_{ij}+f_{ij}(v)+\underbrace{f_{ij}(v+\tilde{\phi_0})-f_{ij}(v)  }_{\tilde{f}_{ij}(v, \tilde{\phi_0})}\right]\,.
\end{aligned}
\end{equation}
By performing field transformations

\begin{equation}
\phi_i = ( \xi , \, \eta) \doteq \mathcal{R}_{ij}(v) \,\tilde{\phi}_{0j} \, ,
\end{equation}
we obtain

\begin{equation}
\begin{aligned}
\frac{1}{2} \partial \phi_{0i} \,\partial \phi_{0j} \left(\delta_{ij}+f_{ij}(\phi_0)\right) &=  \frac{1}{2}\partial \phi_i\partial \phi_j \left[\delta_{ij}+ \underbrace{ \left( \mathcal{R}^{-1T} \tilde{f}(v, \mathcal{R}^{-1} \phi) \mathcal{R}^{-1}\right)_{ij}}_{\bar{f}_{ij}(v, \phi) }   \right] =\frac{1}{2}\partial \phi_i\partial \phi_j \left[\delta_{ij}+\bar{f}_{ij}(v, \phi)\right] ,
\end{aligned}
\end{equation}
where 

\begin{equation}
\bar{f}(v, \phi)= \mathcal{R}^{-1T} \left[f(v+ \mathcal{R}^{-1} \phi)-f(v) \right] \mathcal{R}^{-1}\,.
\end{equation}
Explicitly, we find for $ \delta_{ij}+\bar{f}_{ij}(v, \phi) $:

\begin{equation}
1+\bar{f}_{11} =   \dfrac{1-\frac{\eta^2}{f^2}}{1-\frac{\eta^2}{f^2-v^2-v_\eta^2}-\frac{\xi^2}{f^2}-\frac{2v \xi}{f \sqrt{f^2-v^2-v_\eta^2}}\sqrt{1+v_\eta^2/v^2}} \ ,
\end{equation}

\begin{equation}
1+\bar{f}_{22}=  \dfrac{1-\frac{\xi}{f^2}-\frac{2v \xi}{f\sqrt{f^2-v^2-v_\eta^2}}\sqrt{1+v_\eta^2/v^2}}{1-\frac{\eta^2}{f^2-v^2-v_\eta^2}-\frac{\xi^2}{f^2}-\frac{2v \xi}{f\sqrt{f^2-v^2-v_\eta^2}}\sqrt{1+v_\eta^2/v^2}}  \ ,
\end{equation}

\begin{equation}
\bar{f}_{12}=  \dfrac{- \xi \eta- \frac{fv \eta}{\sqrt{f^2-v^2-v_\eta^2}}\sqrt{1+v_\eta^2/v^2}}{-f^2+\xi^2+\frac{f^2 \eta^2}{f^2-v^2-v_\eta^2}+\frac{2vf \xi}{\sqrt{f^2-v^2-v_\eta^2}}\sqrt{1+v_\eta^2/v^2}} \ .
\end{equation}
Turning to the potential, we see that in order to find the physical fields we have to mass-diagonalize it.
From the form of the potential and (\ref{eqn:vevrelation}) we can deduce the mass mixing matrix, by expanding around the vevs, to be

\begin{equation}
\frac{1}{2} M^2 = \frac{1}{2} \begin{pmatrix}
8 C_1v_1^2 & 4Bv_1v_2 \\ 4 B v_1 v_2& 8 C_2 v_2^2
\end{pmatrix} \ .
\end{equation}
For the canonical fields it becomes

\begin{equation}
\frac{1}{2} \bar{M}^2 = \frac{1}{2} \mathcal{R}^{-1T} \begin{pmatrix}
8 C_1v_1^2 & 4Bv_1v_2 \\ 4 B v_1 v_2& 8 C_2 v_2^2
\end{pmatrix} \mathcal{R}^{-1} \ .
\label{eqn:Mbar}
\end{equation}
We diagonalize it via a diagonalization matrix $ R(\theta) $,

\begin{equation}
R(\theta) = \begin{pmatrix} \cos \theta & \sin \theta \\ - \sin \theta & \cos \theta \end{pmatrix} \, ,
\end{equation}
with the mixing angle $ \theta $, such that

\begin{equation}
\mathcal{R}^{-1T} \begin{pmatrix}
8 C_1v_1^2 & 4Bv_1v_2 \\ 4 B v_1 v_2& 8 C_2 v_2^2
\end{pmatrix} \mathcal{R}^{-1} = R(\theta) \mathcal{M}^2 R^T(\theta) \, ,
\end{equation}
with the diagonal matrix

\begin{equation}
\mathcal{M}^2 = \text{diag} (m^2 , M^2) \, .
\end{equation}
Hence,

\begin{equation}
\begin{aligned}
\begin{pmatrix}
8 C_1v_1^2 & 4Bv_1v_2 \\ 4 B v_1 v_2& 8 C_2 v_2^2
\end{pmatrix} & = \mathcal{R}^T(v) R(\theta) {\mathcal{M}}^2 R^T(\theta) \,\mathcal{R}(v)  \doteq W^{-1T}(\theta, v)\, \mathcal{M}^2 \,W^{-1}(\theta, v) \, ,
\end{aligned}
\end{equation}
with

\begin{equation}
W(\theta, v)= \mathcal{R}^{-1}(v)R(\theta) \, .
\end{equation}
We find

\begin{equation}
\begin{aligned}
W = \dfrac{1}{fv \sqrt{1+v_\eta^2/v^2}} \times \begin{pmatrix}
v \sqrt{f^2-v^2-v_\eta^2} \cos \theta - f v_\eta \sin \theta & f v_\eta \cos \theta + v \sqrt{f^2-v^2-v_\eta^2}\sin \theta \\
v_\eta \sqrt{f^2-v^2-v_\eta^2} \cos \theta + f v \sin \theta & - f v \cos \theta + v_\eta \sqrt{f^2-v^2-v_\eta^2}\sin \theta 
\end{pmatrix}  .
\end{aligned}
\end{equation}
With this, one can express the original model parameters in terms of the new physical parameters $m^2, M^2, v_1, v_2, \theta$:

\begin{equation}\label{eqn:CsB}
\begin{aligned}
C_1& =\frac{1}{8v_1^2} \left(W^{-1T} \mathcal{M}^2 W^{-1}\right)_{11} \, , \\
C_2 & = \frac{1}{8v_2^2} \left(W^{-1T} \mathcal{M}^2 W^{-1}\right)_{22} \, , \\
B &=  \frac{1}{4v_1v_2} \left(W^{-1T} \mathcal{M}^2 W^{-1}\right)_{12} \, .
\end{aligned}
\end{equation}
Inserting into the expressions for $A_1$ and $A_2$ (\ref{eqn:vevrelation}) we get

\begin{equation}\label{eqn:As}
\begin{aligned}
A_1& =-\frac{v_2}{4v_1} \left(W^{-1T} \mathcal{M}^2 W^{-1}\right)_{12}-\frac{1}{4} \left(W^{-1T} \mathcal{M}^2 W^{-1}\right)_{11} \, , \\
A_2& =-\frac{v_1}{4v_2} \left(W^{-1T} \mathcal{M}^2 W^{-1}\right)_{12}-\frac{1}{4} \left(W^{-1T} \mathcal{M}^2 W^{-1}\right)_{22}\,.
\end{aligned}
\end{equation}
For the explicit expressions of $W^{-1T} \mathcal{M}^2 W^{-1}$, see Equations (\ref{eqn:Wexplicit}) below.
Defining the mass eigenstate field by 

\begin{equation}
\Phi =\begin{pmatrix}
h\\ S
\end{pmatrix} = R^T(\theta) \phi =  R^T(\theta) \mathcal{R}(v)\tilde{\phi}_0 \, ,
\end{equation}
and reversed,

\begin{equation}
\tilde{\phi}_0= \mathcal{R}^{-1}(v) R(\theta) \Phi = W(\theta, v) \Phi\, ,
\end{equation}
we find in detail:

\begin{equation}
\begin{aligned}
\begin{pmatrix}
\xi_0 \\ \eta_0
\end{pmatrix}=\begin{pmatrix}
v\\v_\eta
\end{pmatrix}+\mathcal{R}^{-1} R \begin{pmatrix}
h \\ S
\end{pmatrix} = 
\begin{pmatrix}
v+ \left( \frac{u}{fw} \mathbf{c} -\frac{v_\eta}{v w} \mathbf{s}  \right) h + \left( \frac{v_\eta}{v w} \mathbf{c} + \frac{u}{fw} \mathbf{s} \right) S \\v_\eta + \left( \frac{v_\eta u}{fv w} \mathbf{c} + \frac{\mathbf{s}}{w} \right) h + \left( - \frac{\mathbf{c}}{w} + \frac{v\eta u}{fvw} \mathbf{s} \right) S  
\end{pmatrix}
.
\end{aligned}
\label{eqn:originalphysical}
\end{equation}
We used some shorthands that are going to be used frequently in the following to display expressions:

\begin{equation}
u=\sqrt{f^2-v^2-v_\eta^2} \, , \ v_0^2 = v^2+v_\eta^2 \, , \ w = \sqrt{1+\frac{v^2}{v_\eta^2}} \, , \ \cos \theta = \mathbf{c} \, , \ \sin \theta = \mathbf{s} \ .
\label{eqn:shorthands}
\end{equation}
We can thus rewrite the original Lagrangian (\ref{eqn:electroweakSO6}) in terms of the physical fields:

\begin{equation}\label{eqn:finalL}
\begin{aligned}
\mathcal{L} & =-\frac{1}{4} W_{\mu\nu}^{a2}-\frac{1}{4} B_{\mu\nu}^2 + \frac{v_1^2}{4}\langle L_\mu L^\mu\rangle \left( \underbrace{1+\frac{2}{v_1} (W \Phi)_1 +\frac{1}{v_1^2}(W\Phi)_1^2}_{F_U}\right) \\
& + \frac{1}{2}\partial \Phi^T \left[ \underbrace{ \mathbf{1}+W^T\left(f(v+W\Phi)-f(v)\right) W }_{1+F(v,\Phi)}\right] \partial \Phi -V(h,S)  + \bar{\psi} M(v+W\Phi, \, U) \psi\,.
\end{aligned}
\end{equation}
Where the potential $V(h,S)$ reads

\begin{equation}
\begin{aligned}
V(h,S) =& V(v)+\frac{1}{2}\Phi^T \mathcal{M}^2 \Phi
+\frac{1}{3!}\left[V^{(jkl)}_3 W_{jm} W_{kn} W_{lp}\right] \Phi_m \Phi_n \Phi_p \\
& +\frac{1}{4 !}\left[V^{(ijkl)}_4 W_{im} W_{jn} W_{kp} W_{lq}\right] \Phi_m \Phi_n \Phi_p \Phi_q \ .
\end{aligned}
\end{equation}
We directly obtain the function $ F_U(h,S) $ as

\begin{equation}
\begin{aligned}
F_U(h,S) &= 1+ \frac{2}{fv^2w} \big( (vu \mathbf{c} - f v_\eta \mathbf{s})\, h + (f v_\eta \mathbf{c} + vu \mathbf{s}) \, S \big) \\
& \quad + \frac{1}{f^2v^4w^2} \big( (v^2 u^2 \mathbf{c}^2 + f^2 v_\eta^2 \mathbf{s}^2 - 2 fv v_\eta u \mathbf{c} \mathbf{s})\, h^2  \\
& \qquad \qquad + (v^2 u^2 \mathbf{c} \mathbf{s} - f^2 v_\eta^2 \mathbf{c} \mathbf{s} + fvv_\eta u (\mathbf{c}^2-\mathbf{s}^2))\, hS \\
& \qquad \qquad + (f^2 v_\eta^2 \mathbf{c}^2 + v^2 u^2 \mathbf{s}^2 + 2 fvv_\eta u \mathbf{c} \mathbf{s} ) \, S^2 \big)  \ .
\end{aligned}
\label{eqn:FUSO6}
\end{equation}

The potential for the model in terms of the parameter set ($ v,v_\eta, \theta, m ,M $) is now known but very cumbersome.
The effective Yukawa couplings are yet to be measured if the model can be harmonized with nature.

In detail, we find for $ W^{-1T} \mathcal{M}^2 W^{-1} $ in (\ref{eqn:CsB},\ref{eqn:As}), to express the old potential parameters in terms of the new physical set of parameters $m^2, M^2, v_1, v_\eta, \theta$:

\begin{equation}
\begin{aligned}
&(W^{-1T} \mathcal{M}^2 W^{-1})_{11}\\
& =- \left[ (v^2+v_\eta^2) (f^2-v^2-v_\eta^2) \right]^{-1} \Big[ \left( M^2v_\eta^2 (v^2+v_\eta^2)
-f^2(m^2v^2+M^2v_\eta^2) \right) \cos^2(\theta) \\
& +\left( m^2 v_\eta^2(v^2+v_\eta^2)-f^2(M^2v^2+m^2v_\eta^2) \right) \sin^2(\theta) -f(M^2-m^2) v v_\eta \sqrt{f^2-v^2-v_\eta^2} \, \sin (2\theta) \Big] \ , \\ 
&(W^{-1T} \mathcal{M}^2 W^{-1})_{12}\\
& = (W^{-1T} \mathcal{M}^2 W^{-1})_{21} = \left[ (v^2+v_\eta^2) (f^2-v^2-v_\eta^2) \right]^{-1} \Big[ v v_\eta \left( M^2 (v^2+v_\eta^2) - f^2 (M^2-m^2) \right) \cos^2 (\theta) \\
& - f (M^2-m^2) \sqrt{f^2-v^2-v_\eta^2} (v^2-v_\eta^2) \cos (\theta) \sin (\theta)  + v v_\eta \left( m^2 (v^2+v_\eta^2)+ f^2 (M^2-m^2) \right) \sin^2 (\theta)  \Big] \ , \\ 
&(W^{-1T} \mathcal{M}^2 W^{-1})_{22}\\
& = -\left[ (v^2+v_\eta^2) (f^2-v^2-v_\eta^2) \right]^{-1} \Big[ \left( M^2 v^2 (v^2+v_\eta^2) - f^2 (M^2 v^2+m^2 v_\eta^2) \right) \cos^2 (\theta)\\
&+ \left( m^2 v^2 (v^2 +v_\eta^2) - f^2 (M^2 v_\eta^2+m^2 v^2) \right) \sin^2 (\theta)  + f (M^2-m^2) v v_\eta \sqrt{f^2-v^2-v_\eta^2} \, \sin (2 \theta) \Big] \ .
\end{aligned}
\label{eqn:Wexplicit}
\end{equation}
The kinetic terms expressed via the physical fields in (\ref{eqn:finalL}) are

{\small
	\begin{equation}
	\begin{aligned}
	& 1+ F_h \doteq  1+ F_{11} \\
	&=\dfrac{(f^2-S^2) u^2-2fSvuw \sin(\theta)}{f^2u^2- (h^2+S^2) u^2 -2fv (h \cos (\theta) + S \sin(\theta)) uw + \frac{1}{2}(h^2-S^2) (v^2+v_\eta^2) \cos(2\theta) + hS (v^2+v_\eta^2) \sin(2\theta)} ,
	\end{aligned}
	\end{equation}

	\begin{equation}
	\begin{aligned}
	& 1+ F_S \doteq 1+ F_{22} \\
	&=\dfrac{(f^2-h^2) u^2-2fhvuw \cos(\theta)}{f^2u^2- (h^2+S^2) u^2 -2fv (h \cos (\theta) + S \sin(\theta)) uw + \frac{1}{2}(h^2-S^2) (v^2+v_\eta^2) \cos(2\theta) + hS (v^2+v_\eta^2) \sin(2\theta)}  ,
	\end{aligned}
	\end{equation}

	\begin{equation}
	\begin{aligned}
	& F_{hS} \doteq F_{12}  \\
	& =\dfrac{hS u^2 + fv (h \sin (\theta) + S \cos(\theta)) uw}{f^2 u^2 -2fv (h \cos (\theta) + S \sin (\theta) ) uw - f^2 (h^2+S^2) + (h^2 \cos^2(\theta) + S^2 \sin^2 (\theta)) (v^2+ v_\eta^2) + hS (v^2+v_\eta^2) \sin (2\theta)}  .
	\end{aligned}
	\end{equation}
}

We thus have derived the functions $F_U , \, F_h, \, F_S, \, F_{hS}, \ \textrm{and} \ V $ that have to be put into our formula for the one-scalar-loop divergence derived in \cite{lindnermuzakka2022}.
For the two Higgs fields in the case here the polynomial functions supplementing the kinetic terms $F_h, \, F_S, \, F_{hS}$, are to be put into the corresponding formula to extract the matrices $X$ and $Y$ \eqref{eqn:XandY} and then the master formula \eqref{eqn:tHooft}.
Otherwise, the principal form of extracting the divergences remains unchanged to that of the common electroweak chiral Lagrangian \cite{completerenorm}.

\subsection{Unitarity}
\label{SO6unitarity}

We consider $ W^+W^- \rightarrow W^+W^- $ scattering at high energies -- the corresponding Goldstone scattering.

\begin{figure}[ht]
    \centering
    \includegraphics[width=0.3\textwidth]{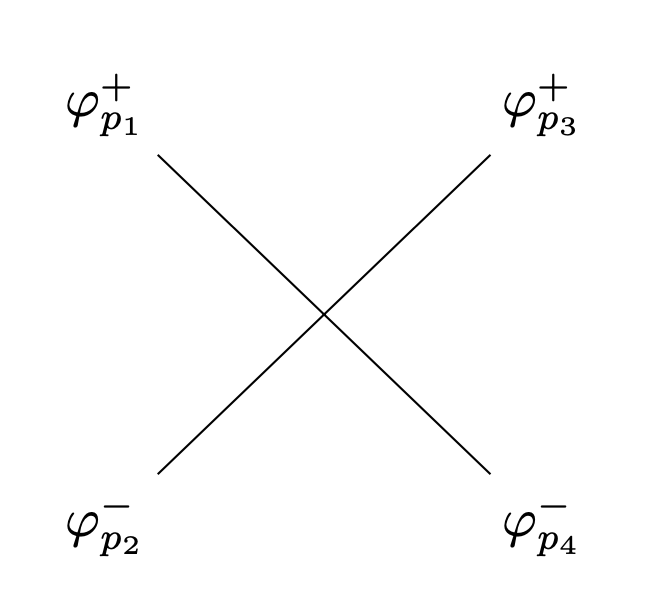}
    \caption{Pure Goldstone scattering}
    \label{fig:phi_scat}
\end{figure}

\begin{figure}[ht]
    \centering
    \includegraphics[width=0.5\textwidth]{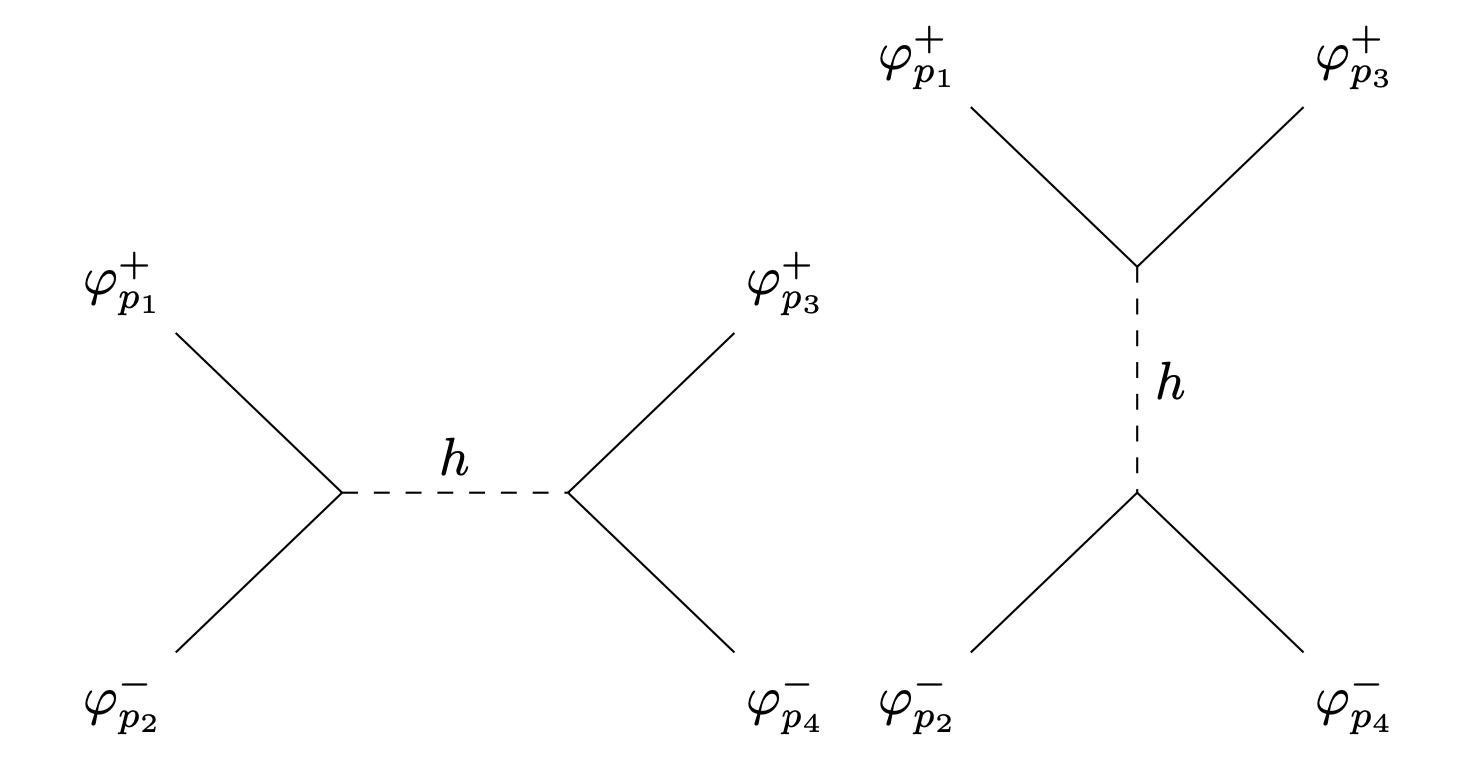}%
    \includegraphics[width=0.5\textwidth]{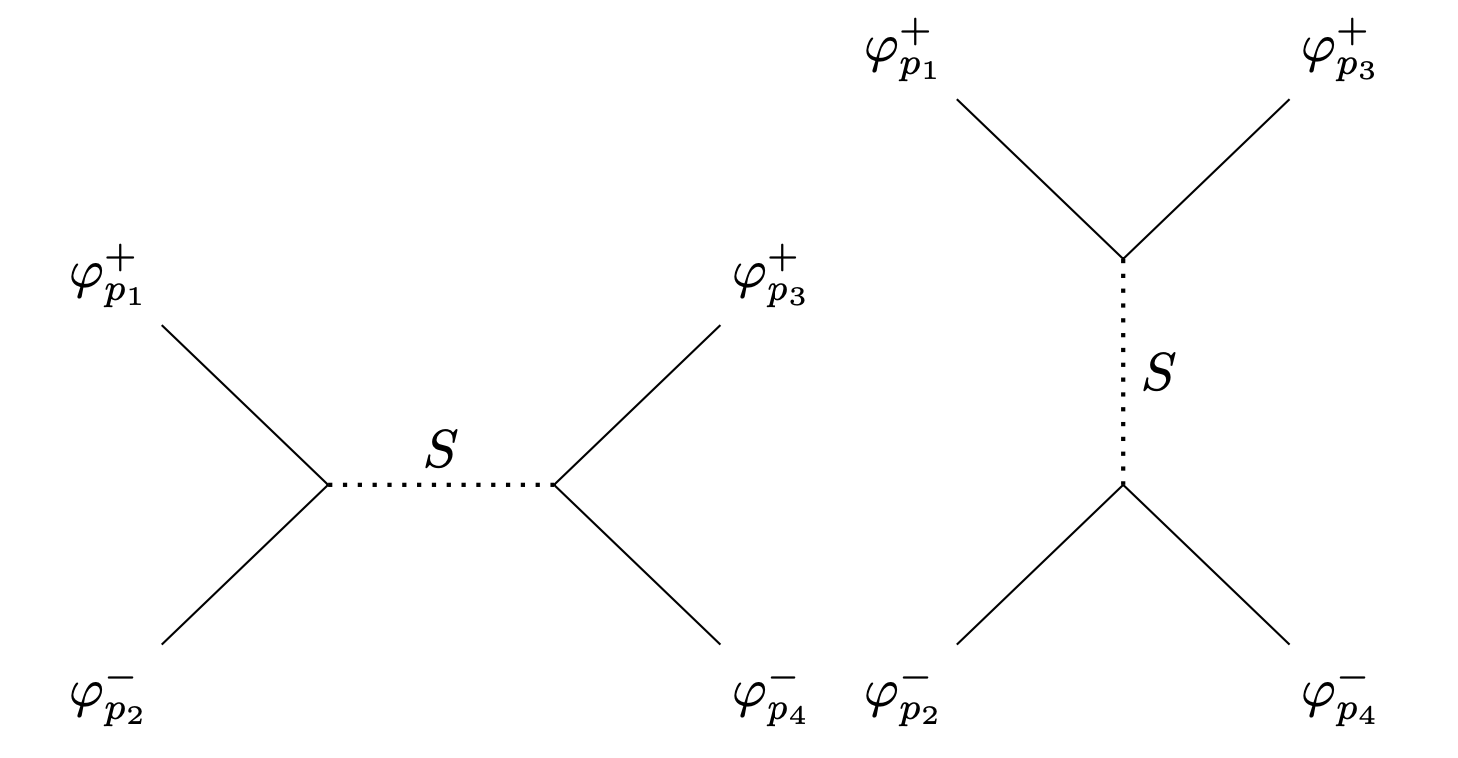}
    \caption{Goldstone scattering mediated by $h$ and $S$}
    \label{fig:phi_scat_h_S}
\end{figure}

In general, for the electroweak chiral Lagrangian with two scalars, we write the Goldstone kinetic term as (with the components of $F_U$ written $F_{ij}$)

\begin{equation}
\mathcal{L}_{\text{G,kin}} = \frac{v^2}{4} \langle D_\mu U^\dagger D^\mu U \rangle \left( F_{10} \frac{h}{v} + F_{01} \frac{S}{v} + F_{11} \frac{hS}{v^2} + F_{20} \frac{h^2}{v^2}+ F_{02} \frac{S^2}{v^2} + ...\right) \ .
\label{eqn:goldscalar}
\end{equation}
For the $ SO(6)/SO(5) $ CHM there are no higher order terms, see Equation (\ref{eqn:FUSO6}).
In terms of the physical Goldstones to second order we have

\begin{equation}
\frac{v^2}{4} \langle D_\mu U^\dagger D^\mu U \rangle \, F_U = ( \partial_\mu \varphi^+ \partial^\mu \varphi^- + \frac{1}{2} \partial_\mu \varphi^0 \partial^\mu \varphi^0 + ... ) \, F_U \ .
\end{equation}
We then find for the contributing diagrams in Figure \ref{fig:phi_scat} and Figure \ref{fig:phi_scat_h_S}


\ifx false
\begin{figure}
	\begin{tikzpicture}
	\begin{feynman}
	\vertex (a);
	\vertex [above left=of a] (p1) {\( \varphi^+_{p_1} \)};
	\vertex [below left=of a] (p2) {\( \varphi^-_{p_2} \)};
	\vertex [above right=of a] (f1) {\( \varphi^+_{p_3} \)};
	\vertex [below right=of a] (f2) {\( \varphi^-_{p_4} \)};

	\diagram*{
		(a) -- (p1);
		(a) -- (p2);
		(a) --  (f1);
		(a) --  (f2);
		
	};
	
	\end{feynman}
	\end{tikzpicture}
	\\
	\begin{tikzpicture}
	\begin{feynman}
	\vertex (a);
	\vertex (b);
	\vertex [above left=of a] (p1) {\( \varphi^+_{p_1} \)};
	\vertex [below left=of a] (p2) {\( \varphi^-_{p_2} \)};
	\vertex [right =1.5cm of a] (b) ;
	\vertex [above right=of b] (f1) {\( \varphi^+_{p_3} \)};
	\vertex [below right=of b] (f2) {\( \varphi^-_{p_4} \)};

	\diagram*{
		(a) -- (p1);
		(a) -- (p2);
		(b) --  (f1);
		(b) --  (f2);
		(a) -- [scalar, edge label = \( h \)] (b) ;
	};
	
	\end{feynman}
	\end{tikzpicture}
	\begin{tikzpicture}
	\begin{feynman}
	\vertex (a);
	\vertex (b);
	\vertex [above left=of a] (p1) {\( \varphi^+_{p_1} \)};
	\vertex [above right=of a] (p2) {\( \varphi^+_{p_3} \)};
	\vertex [below =1.5cm of a] (b) ;
	\vertex [below left =of b] (f1) {\( \varphi^-_{p_2} \)};
	\vertex [below right=of b] (f2) {\( \varphi^-_{p_4} \)};

	\diagram*{
		(a) -- (p1);
		(a) -- (p2);
		(b) --  (f1);
		(b) --  (f2);
		(a) -- [scalar, edge label = \( h \)] (b) ;
	};
	
	\end{feynman}
	\end{tikzpicture}\\
	\begin{tikzpicture}
	\begin{feynman}
	\vertex (a);
	\vertex (b);
	\vertex [above left=of a] (p1) {\( \varphi^+_{p_1} \)};
	\vertex [below left=of a] (p2) {\( \varphi^-_{p_2} \)};
	\vertex [right =1.5cm of a] (b) ;
	\vertex [above right=of b] (f1) {\( \varphi^+_{p_3} \)};
	\vertex [below right=of b] (f2) {\( \varphi^-_{p_4} \)};

	\diagram*{
		(a) -- (p1);
		(a) -- (p2);
		(b) --  (f1);
		(b) --  (f2);
		(a) -- [ghost, edge label = \( S \)] (b) ;
	};
	
	\end{feynman}
	\end{tikzpicture}
	\begin{tikzpicture}
	\begin{feynman}
	\vertex (a);
	\vertex (b);
	\vertex [above left=of a] (p1) {\( \varphi^+_{p_1} \)};
	\vertex [above right=of a] (p2) {\( \varphi^+_{p_3} \)};
	\vertex [below =1.5cm of a] (b) ;
	\vertex [below left =of b] (f1) {\( \varphi^-_{p_2} \)};
	\vertex [below right=of b] (f2) {\( \varphi^-_{p_4} \)};

	\diagram*{
		(a) -- (p1);
		(a) -- (p2);
		(b) --  (f1);
		(b) --  (f2);
		(a) -- [ghost, edge label = \( S \)] (b) ;
	};
	
	\end{feynman}
	\end{tikzpicture}
\end{figure}
\fi

\begin{equation}
\begin{aligned}
& \mathcal{M}( \varphi^+\varphi^- \rightarrow \varphi^+\varphi^-) \\
&= \frac{1}{v^2}(s+t) \\
&+ \frac{iF_{10}}{v} (-ip_1)(-ip_2) \frac{1}{s-m^2}\frac{iF_{10}}{v} (ip_3)(ip_4) + \frac{iF_{10}}{v} (-ip_1)(-ip_3) \frac{1}{t-m^2}\frac{iF_{10}}{v} (ip_2)(ip_4) 
\\ 
&+ \frac{iF_{01}}{v} (-ip_1)(-ip_2) \frac{1}{s-M^2}\frac{iF_{01}}{v} (ip_3)(ip_4) + \frac{iF_{01}}{v} (-ip_1)(-ip_3) \frac{1}{t-M^2}\frac{iF_{01}}{v} (ip_2)(ip_4) \\
&= \frac{1}{v^2} \Big[ s+t  - \left( \frac{F_{10}}{2} \right)^2 \frac{s^2}{s-m^2} - \left( \frac{F_{10}}{2} \right)^2 \frac{t^2}{t-m^2} - \left( \frac{F_{01}}{2} \right)^2 \frac{s^2}{s-M^2} - \left( \frac{F_{01}}{2} \right)^2 \frac{t^2}{t-M^2} \Big] \\
& = \frac{s+t}{v^2} \left[ 1 - \frac{F_{10}^2+F_{01}^2}{4} \right] +\mathcal{O}(\frac{\text{mass}^2}{E^2}) \ .
\end{aligned}
\end{equation}
We identify the renormalizability constraint for the electroweak chiral Lagrangian with two singlet scalars.
In the renormalizable limit, the vector boson scattering is unitarized.
In this model, inserting the functions $F_{10}$, $F_{01}$ from \eqref{eqn:FUSO6}, the constraint is

\begin{equation}
1- \dfrac{(vu\mathbf{c}-fv_\eta \mathbf{s})^2+(fv_\eta \mathbf{c}+vu \mathbf{s})^2}{f^2 v^2 w^2} \overset{!}{=} 0 .
\end{equation}
We can illustrate the (partly) unitarizing effect of the model by setting $ v_\eta=0 $.
The amplitudes then scale as

\begin{equation}
\mathcal{M} \sim \frac{s}{v^2} \frac{v^2}{f^2} \ .
\end{equation}
Hence, violation of perturbative unitarity is postponed relative to the electroweak chiral Lagrangian from $ \Lambda = 4\pi v $ to $ \Lambda = 4\pi f $.
As we mentioned in the introduction, this is typical for composite Higgs models.

Interestingly, the scattering of Goldstones into scalars is trivially unitarized in the model.
The amplitudes for $ \varphi^+ \varphi^- \rightarrow hh  $ and $ \varphi^+ \varphi^- \rightarrow SS $ scattering, where the $ u $-channel is implicit in the diagrams in Figure \ref{fig:phi_h_S_scat} ($ s $-channel diagrams with an $ h $ or $ S $ propagator are of the order $ \text{mass}^2/E^2 $) are computed to

\begin{figure}[ht]
    \centering
    \includegraphics[width=0.4\textwidth]{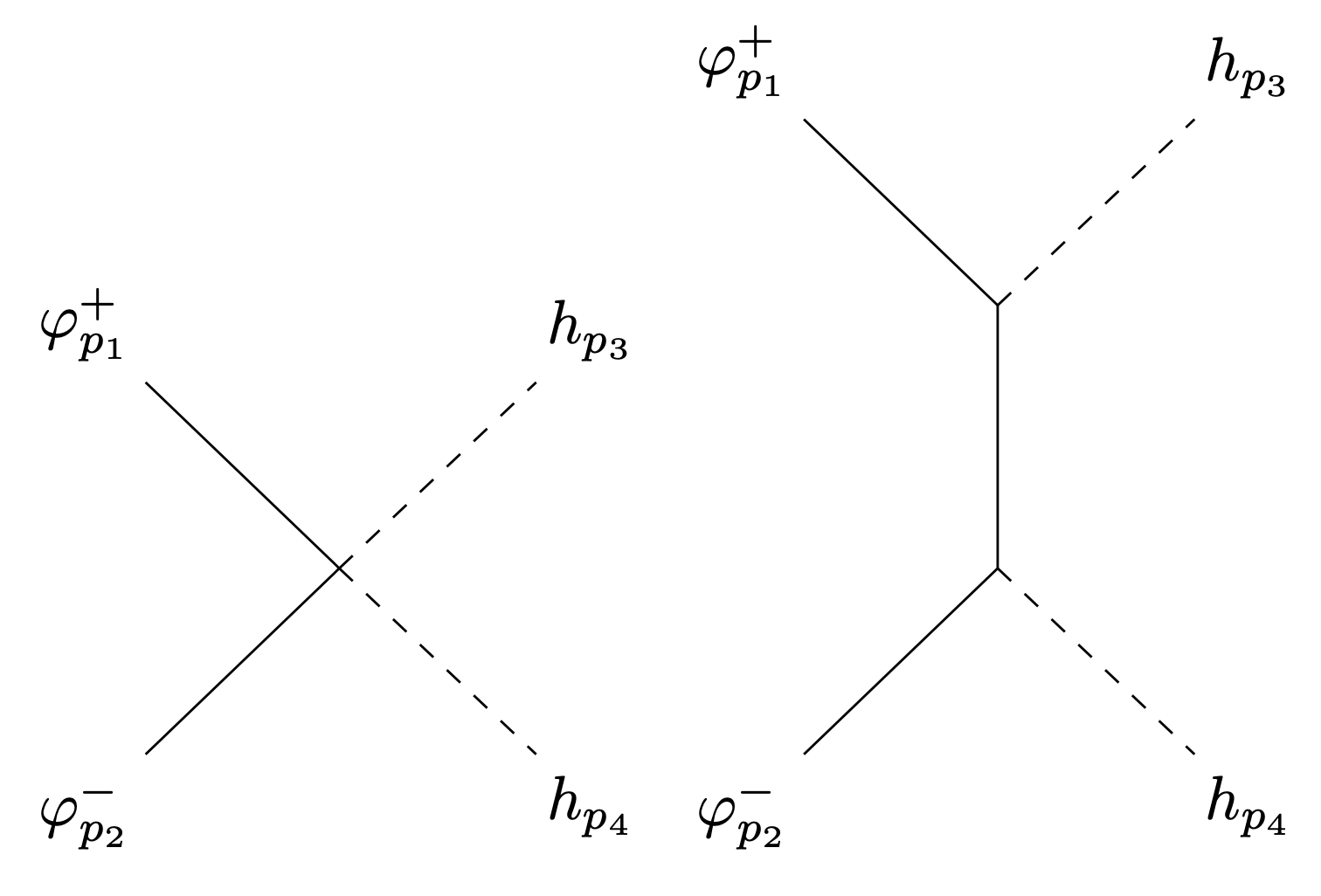}%
    \includegraphics[width=0.4\textwidth]{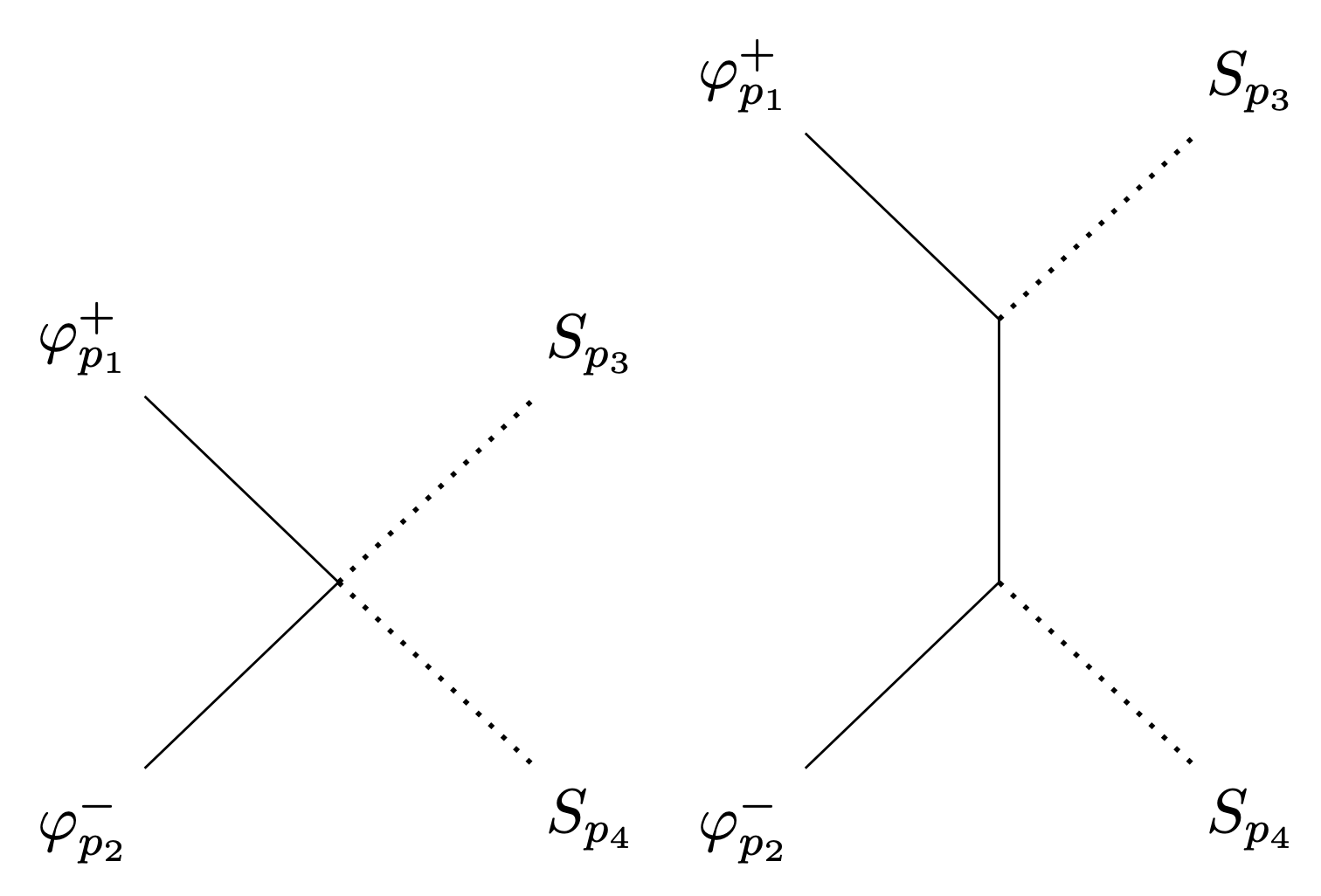}
    \caption{Goldstone into equal singlets scattering}
    \label{fig:phi_h_S_scat}
\end{figure}

\ifx false
\begin{figure}[ht]
	\centering
	\begin{tikzpicture}
	\begin{feynman}
	\vertex (a);
	\vertex [above left=of a] (p1) {\( \varphi^+_{p_1} \)};
	\vertex [below left=of a] (p2) {\( \varphi^-_{p_2} \)};
	\vertex [above right=of a] (f1) {\( h_{p_3} \)};
	\vertex [below right=of a] (f2) {\( h_{p_4} \)};

	\diagram*{
		(a) -- (p1);
		(a) -- (p2);
		(a) -- [scalar]  (f1);
		(a) -- [scalar] (f2);
		
	};
	
	\end{feynman}
	\end{tikzpicture}
	\begin{tikzpicture}
	\begin{feynman}
	\vertex (a);
	\vertex (b);
	\vertex [above left=of a] (p1) {\( \varphi^+_{p_1} \)};
	\vertex [above right=of a] (p2) {\( h_{p_3} \)};
	\vertex [below =1.5cm of a] (b) ;
	\vertex [below left =of b] (f1) {\( \varphi^-_{p_2} \)};
	\vertex [below right=of b] (f2) {\( h_{p_4} \)};

	\diagram*{
		(a) -- (p1);
		(a) -- [scalar] (p2);
		(b) -- (f1);
		(b) -- [scalar] (f2);
		(a) --(b) ;
	};
	
	\end{feynman}
	\end{tikzpicture}
	\begin{tikzpicture}
	\begin{feynman}
	\vertex (a);
	\vertex [above left=of a] (p1) {\( \varphi^+_{p_1} \)};
	\vertex [below left=of a] (p2) {\( \varphi^-_{p_2} \)};
	\vertex [above right=of a] (f1) {\( S_{p_3} \)};
	\vertex [below right=of a] (f2) {\( S_{p_4} \)};

	\diagram*{
		(a) -- (p1);
		(a) -- (p2);
		(a) -- [ghost] (f1);
		(a) -- [ghost] (f2);
		
	};
	
	\end{feynman}
	\end{tikzpicture}
	\begin{tikzpicture}
	\begin{feynman}
	\vertex (a);
	\vertex (b);
	\vertex [above left=of a] (p1) {\( \varphi^+_{p_1} \)};
	\vertex [above right=of a] (p2) {\( S_{p_3} \)};
	\vertex [below =1.5cm of a] (b) ;
	\vertex [below left =of b] (f1) {\( \varphi^-_{p_2} \)};
	\vertex [below right=of b] (f2) {\( S_{p_4} \)};

	\diagram*{
		(a) -- (p1);
		(a) -- [ghost] (p2);
		(b) -- (f1);
		(b) -- [ghost](f2);
		(a) --(b) ;
	};
	
	\end{feynman}
	\end{tikzpicture}
    \caption{Goldstone into equal singlets scattering}
    \label{fig:phi_h_S_scat}
\end{figure}
\fi

\begin{equation}
\begin{aligned}
&\mathcal{M}(\varphi^+\varphi^- \rightarrow hh) \\
&= \frac{2F_{20}}{v^2} (-ip_1)(-ip_2) \\
&+ \frac{iF_{10}}{v} (-ip_{1\mu}) \frac{(p_1-p_3)^\mu (p_1-p_3)^\nu}{t} (-ip_{2\nu}) \frac{iF_{10}}{v} + \frac{iF_{10}}{v} (-ip_{1\mu}) \frac{(p_1-p_4)^\mu (p_1-p_4)^\nu}{u} (-ip_{2\nu}) \frac{iF_{10}}{v} \\
&= -\frac{F_{20}}{v^2} s +\left( \frac{F_{10}}{2v}\right) ^2 (s+u-m^2) +\left( \frac{F_{10}}{2v}\right) ^2 (s+t-m^2) = \frac{s}{v^2} \left[ \left( \frac{F_{10}}{2}\right) ^2- F_{20}  \right] \ , \\
&\mathcal{M}(\varphi^+\varphi^- \rightarrow SS) = \frac{s}{v^2} \left[ \left( \frac{F_{01}}{2}\right) ^2 - F_{02}  \right] \ .
\end{aligned}
\end{equation}
and for $ \varphi^+ \varphi^- \rightarrow hS $ scattering analogously, but without a symmetry factor of 2 in the first diagram in Figure \ref{fig:phi_hS_scat}

\begin{figure}[ht]
    \centering
    \includegraphics[width=0.4\textwidth]{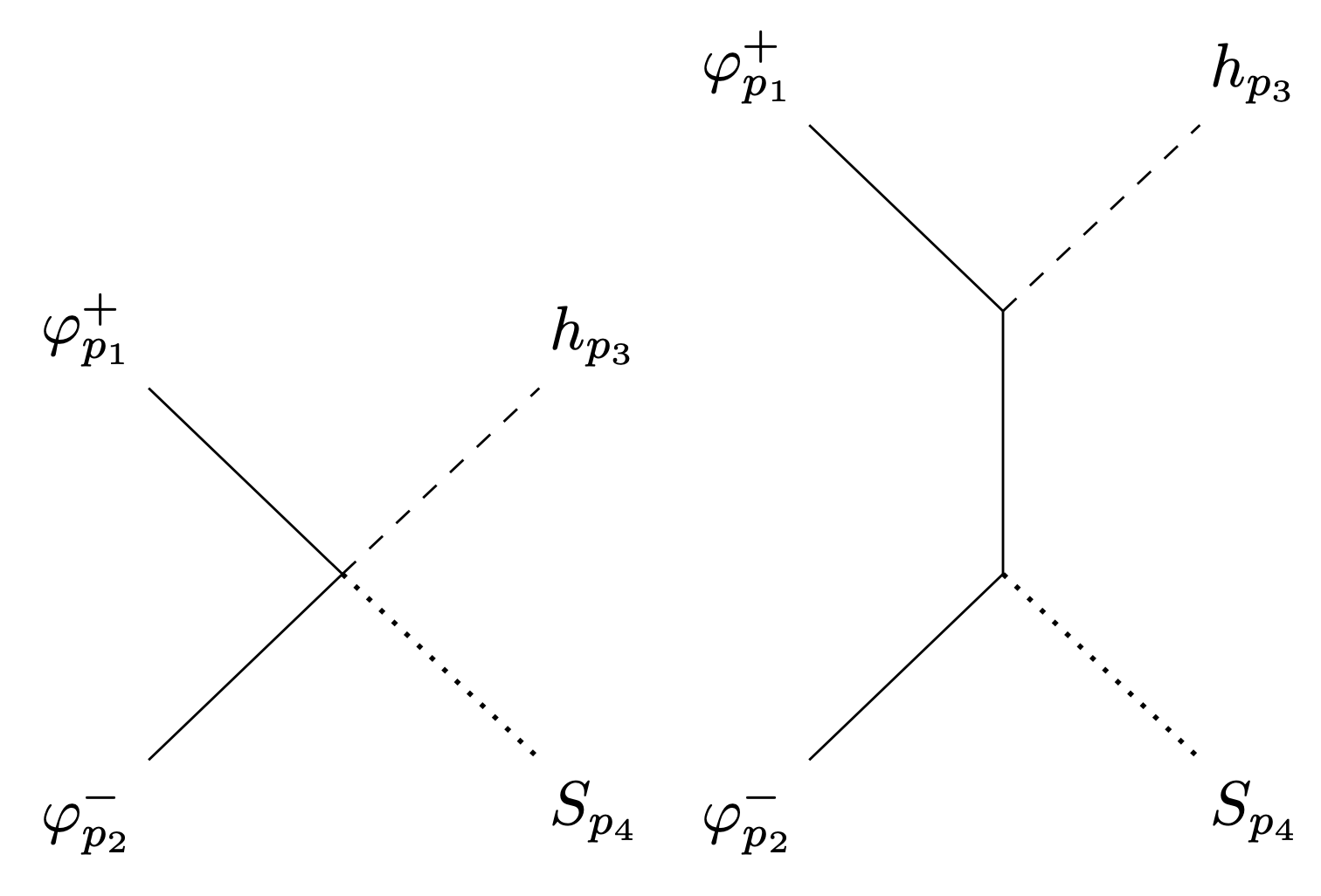}
    \caption{Goldstone into different singlets scattering}
    \label{fig:phi_hS_scat}
\end{figure}

\ifx false
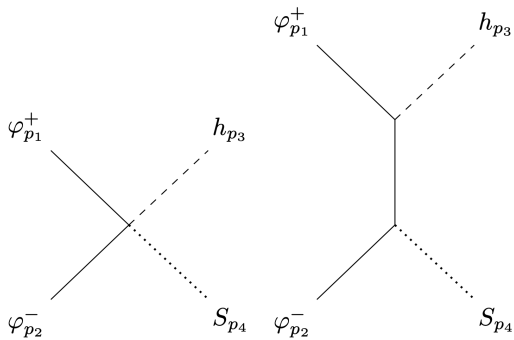
\begin{figure}[ht]
	\centering
	\begin{tikzpicture}
	\begin{feynman}
	\vertex (a);
	\vertex [above left=of a] (p1) {\( \varphi^+_{p_1} \)};
	\vertex [below left=of a] (p2) {\( \varphi^-_{p_2} \)};
	\vertex [above right=of a] (f1) {\( h_{p_3} \)};
	\vertex [below right=of a] (f2) {\( S_{p_4} \)};

	\diagram*{
		(a) -- (p1);
		(a) -- (p2);
		(a) -- [scalar] (f1);
		(a) -- [ghost] (f2);
		
	};
	
	\end{feynman}
	\end{tikzpicture}
	\begin{tikzpicture}
	\begin{feynman}
	\vertex (a);
	\vertex (b);
	\vertex [above left=of a] (p1) {\( \varphi^+_{p_1} \)};
	\vertex [above right=of a] (p2) {\( h_{p_3} \)};
	\vertex [below =1.5cm of a] (b) ;
	\vertex [below left =of b] (f1) {\( \varphi^-_{p_2} \)};
	\vertex [below right=of b] (f2) {\( S_{p_4} \)};

	\diagram*{
		(a) -- (p1);
		(a) -- [scalar] (p2);
		(b) -- (f1);
		(b) -- [ghost](f2);
		(a) --(b) ;
	};
	
	\end{feynman}
	\end{tikzpicture}
    \caption{Goldstone into different singlets scattering}
    \label{fig:phi_hS_scat}
\end{figure}
\fi

\begin{equation}
\mathcal{M}(\varphi^+\varphi^- \rightarrow hS) = \frac{s}{v^2} \left[ \frac{F_{10}F_{01}}{4}-\frac{F_{11}}{2}  \right] \ .
\end{equation}
For the parameters of this model, all these contributions cancel each other.

\subsection{Integrating out the Heavy Scalar -- Matching to the Electroweak Chiral Lagrangian with one light Higgs}
\label{integrateout}

To match the model to the EWCLh we have to assume $ S $ to be substantially heavier than $ h $, such that we can integrate it out alone.
We can then make the immediate connection to the current situation of one observed Higgs-like scalar and predict the values of its effective couplings.
It is a natural guess to assume the mass of $ S $ to be in the range of 500 GeV.
As a pseudo-Goldstone it is naturally permitted to be lighter than the scale $ f $ that is assumed to be at around 1 TeV.
\\
Leaving aside the gauge-kinetic terms the $SO(6)/SO(5)$ composite Higgs model is given by 

\begin{equation}
\begin{aligned}
\mathcal{L} & = \frac{v_1^2}{4}\langle L_\mu L^\mu\rangle \left(1+\frac{2}{v_1} (W \Phi)_1+\frac{1}{v_1^2}(W\Phi)_1^2\right) + \frac{1}{2}\partial \Phi^T \left[\mathbf{1}+W^T\left(f(v+W\Phi)-f(v)\right) W\right] \partial \Phi\\
& - \underbrace{ \Bigg[V(v)+\frac{1}{2}\Phi^T \mathcal{M}^2 \Phi
+\frac{1}{3!}\left[V^{(jkl)}_3 W_{jm} W_{kn} W_{lp}\right] \Phi_m \Phi_n \Phi_p  +\frac{1}{4 !}\left[V^{(ijkl)}_4 W_{im} W_{jn} W_{kp} W_{lq}\right] \Phi_m \Phi_n \Phi_p \Phi_q\Bigg] }_{V(h, S)}\\
&+ \bar{\psi} M(v+W\Phi, \, U) \psi\,.
\end{aligned}
\end{equation}
We compactify it:

\begin{equation}
\begin{aligned}
\mathcal{L} & = \frac{1}{2}(1+F_S(h,S)) (\partial S)^2+ F_{hS}(h, S) \partial h\, \partial S +  \frac{1}{2}(1+F_h(h,S)) (\partial h)^2 \\
&- V(h,S)+\frac{v_1^2}{4}\langle L_\mu L^\mu\rangle \left(1+F_U(h, S)\right) + \bar{\psi} M(v+W\Phi, \, U) \psi\\
& \doteq \frac{1}{2} \left(1+F_S\right)(\partial S)^2+ F_{hS}\,\partial h \,\partial S + \mathcal{V}(h, S) \,,
\end{aligned}\label{hdd}
\end{equation}
where we defined

\begin{equation}
\begin{aligned}
\mathcal{V}(h, S) & =  \frac{1}{2}(1+F_h(h,S)) (\partial h)^2- V(h,S)+\frac{v_1^2}{4}\langle L_\mu L^\mu\rangle \left(1+F_U(h, S)\right) + \bar{\psi} M(v+W\Phi, \, U) \psi 
\\
& \doteq \sum _a J_a(h) \,S^a\,.
\end{aligned}
\end{equation}
Our aim is to integrate out the $S$ field under the assumption $M \gg m$ in the strongly coupled limit $ v \lesssim f $ \cite{addsinglet}.
Due to the mass diagonalization procedure we carried out before, we find that the potential parameters functions of the particle masses, mixing angles and vevs.
Hence, we may write

\begin{equation}
\begin{aligned}
\mathcal{V}(h, S) & = \sum_a J_a(h) \,S^a = \sum _a \left[ M^2 J_a^0(h)+\bar{J}_a(h)\right] S^a \, .
\end{aligned}
\label{eqn:decomposition}
\end{equation}
To integrate out the heavy scalar $S$ at tree level, we write $S$ as power series in $M^{-2}$,

\begin{equation}
S=\sum_{n=0} S_n(h) M^{-2n} \, ,
\end{equation}
and insert the whole expansion into its classical equation of motion (E.O.M.) to find the coefficients of the power series.
One can directly compute the equation of motion of $S$ from the Lagrangian

\begin{equation}
\begin{aligned}
\frac{\delta L}{\delta S} &=-\left(1+F_S\right)\partial^2 S-\frac{1}{2} \frac{\partial F_S}{\partial S}(\partial S)^2-\frac{\partial F_S}{\partial h}\partial h\,\partial S  +\frac{\partial \mathcal{V}}{\partial S} -F_{hS}\partial^2 h-\frac{\partial F_{hS}}{\partial h}(\partial h)^2 =0 \ .
\end{aligned}
\end{equation}
To obtain $S_n$ for a general n, we have to derive a recursive formula from the E.O.M.
We introduce the following notation for arbitrary functions $F(h, S)$:

\begin{equation}
F(h, S) = F^{(a)}(h) \,S^a\,,\quad\quad  F^{(a)}(h)= \frac{1}{a!}\frac{\partial^a F }{(\partial S)^a}.
\end{equation}
We then find

\begin{equation}
\begin{aligned}
-\left(1+F_S\right)\partial^2  S & =- \partial^2 S_n M^{-2n}-F_S^{(a)}S_{b_1}S_{b_2}...S_{b_a}\, \partial^2 S_n \, M^{-2(n+b_1+...+b_a)} \ , \\
-\frac{1}{2} \frac{\partial F_S}{\partial S}(\partial S)^2 & =-\frac{1}{2} a F_S^{(a)} S_{b_1}...S_{b_{a-1}} \partial S_m\, \partial S_p\, M^{-2\left[m+p+b_1+...b_{a-1}\right]} \ , \\
-\frac{\partial F_S}{\partial h}\partial h\,\partial S &= -\left(\frac{\partial F_S}{\partial h}\right)^{(a)} S_{b_1}...S_{b_a} \,\partial h\,\partial S_n \,M^{-2(n+b_1+...+b_a)} \ , \\
\frac{\partial \mathcal{V}}{\partial S} -F_{hS}\partial^2 h-\frac{\partial F_{hS}}{\partial h}(\partial h)^2  & = a\,J_a^0 S_{b_1}...S_{b_{a-1}} M^{-2[b_1+...+b_{a-1}-1]}  +a\bar{J}_a S_{b_1}...S_{b_{a-1}} M^{-2(b_1+...+b_{a-1})} \\
& \quad -\left[f_{hS}\partial^2 h+\frac{\partial F_{hS}}{\partial h}(\partial h)^2\right]^{(a)} S_{b_1}...S_{b_{a}} M^{-2(b_1+...+b_a)} \ .
\end{aligned}
\end{equation}
We obtain a recursion relation:

\begin{equation}
\begin{aligned}
0 &=-\partial^2 S_r -\sum_{n+b_1+...+b_a=r}\left[F_S^{(a)} \partial^2S_n+ \left(\frac{\partial F_S}{\partial h}\right)^{(a)} \,\partial h\,\partial S_n \right] S_{b_1}...S_{b_a}   \\
&   -\sum_{m+p+b_1+...b_{a-1}=r}\frac{1}{2} a F_S^{(a)} S_{b_1}...S_{b_{a-1}}\partial S_m\, \partial S_p \\
& +\sum_{b_1+...+b_{a-1}-1=r}a\,J_a^0 S_{b_1}...S_{b_{a-1}}+\sum_{b_1+...+b_{a-1}=r}a\bar{J}_a S_{b_1}...S_{b_{a-1}}\\
&-\sum_{b_1+...+b_{a}=r}\left[F_{hS}\partial^2 h+\frac{\partial F_{hS}}{\partial h}(\partial h)^2\right]^{(a)} S_{b_1}...S_{b_a}\ . \label{qdd}
\end{aligned}
\end{equation}
For instance, lets take $r=-1$; then only the fourth term on the RHS survives, which takes the index values $b_1=...=b_{a-1}=0$.
Therefore, we have 

\begin{equation}
\sum_{a=1}^4 a \,J_a^0 \,S_0^{a-1} =0 \ , 
\label{r-1}
\end{equation}
which can theoretically be solved to get $S_0$.
Setting $r=0$, we get

\begin{equation}
\begin{aligned}
0 & =-\partial^2 S_0  -\sum_{a}\left[F_S^{(a)}\, \partial^2 S_0 +\left(\frac{\partial F_S}{\partial h}\right)^{(a)} \,\partial h\,\partial S_0 \right] S^{a }_0 \\
& -\frac{1}{2}( \partial S_0)^2\sum_{a} a \,F_S^{(a)} S^{a-1}_0	+S_1\sum_a a(a-1)\,J_a^0 S_0^{a-2} +\sum_{a}a\bar{J}_a S_0^{a-1}  \\ 
&-\sum_{a}\left[F_{hS}\partial^2 h+\frac{\partial F_{hS}}{\partial h}(\partial h)^2\right]^{(a)} S_0^a  \ .
\end{aligned}
\end{equation}
Solving this equation gives us $S_1(h)$.
The other terms $S_n$ can be obtained recursively using (\ref{qdd}).
Lets assume that the solution of (\ref{r-1}) for $S_0(h)$ is well defined in the limit $h=0$.
$S_0$ can then be expanded in a power series of $h$ as

\begin{equation}\label{120}
S_0 = \sum_{k=2} r_{k}h^k \ .
\end{equation}
Having obtained $S_0$, one can extract $S_1$ from the recursion relation as a function of $h$.
Thus, the E.O.M. gives us

\begin{equation}
S(h) \doteq S_0(h)+S_1(h)/M^2+... \, .
\end{equation}
We insert this back to the original Lagrangian (\ref{hdd}) and the result is the effective Lagrangian at tree level, organized by a power series expansion in $1/M^2$,

\begin{equation}
\mathcal{L}_{\text{eff}}[h] \doteq \mathcal{L}[h, S_0+1/M^2 S_1+...] = M^2 \mathcal{L}^{(-1)}_{\text{eff}}[h]+\mathcal{L}_{\text{eff}}^{(0)}[h]+\frac{1}{M^2} \mathcal{L}_{\text{eff}}^{(1)}[h]+...
\end{equation}
Now, we prove the following claims:

\begin{enumerate}
\item $S_a(h)$ with $ a \geq 1$ do not contribute to $\mathcal{L}_{\text{eff}}^{(-1)}$ and $\mathcal{L}_{\text{eff}}^{(0)}$.
\item $\mathcal{L}^{(-1)}_{\text{eff}}$ is independent of $h$.
\end{enumerate}
\ \\
To point 1.\\
The original Lagrangian can be written as 

\begin{equation}\label{122}
\mathcal{L}[h, S] = M^2 \mathcal{L}^{(-1)}(h, S)+\mathcal{L}^{(0)} (h, S) \, .
\end{equation}
Now, by inserting the E.O.M. solution of $S(h)$ into the original Lagrangian, we have  

\begin{equation}
\begin{aligned}
& \mathcal{L}[h, S_0+S_1/M^2+...] = \mathcal{L}[h, S_0]+\frac{\delta \mathcal{L}}{\delta S}\Bigg|_{S_0} \left(S_1/M^2+ S_2/ M^4+...\right)+...\\
& =  M^2 \mathcal{L}^{(-1)}(h, S_0)+\mathcal{L}^{(0)} (h, S_0)+\frac{\delta \mathcal{L}}{\delta S}\Bigg|_{S_0} \left(S_1/M^2+ S_2/ M^4+...\right)+... \, .
\end{aligned}
\end{equation}
To prove claim 1, we need to prove that the $\delta \mathcal{L}/\delta S |_{S_0}$ expansion in $M^2$ starts at $\mathcal{O}(1)$.
This is indeed the case.
We have

\begin{equation}
\frac{\delta \mathcal{L}[h, S]}{\delta S}\Bigg|_{\bar{S} = S_0+S_1/M^2+...} = M^2 E^{(-1)}+ E^ {(0)}+\frac{1}{M^2} E^{(1)}+... =0\,,
\end{equation}
where $E^{(r)}$ is precisely the RHS of (\ref{qdd}).
We obtain

\begin{equation}
\frac{\delta \mathcal{L}}{\delta S}\Bigg|_{S_0}=M^2\sum_a a J^0_a S^{a-1}_0 + E^{(0)}(S_0, S_{a\geq 1}=0)+... =  E^{(0)}(S_0, S_{a\geq 1}=0)+... \, ,
\end{equation} 
where (\ref{r-1}) has been used.
Since  $\delta \mathcal{L}/\delta S |_{S_0}$ begins at $\mathcal{O}(1)$, the claim is validated.
\\

\noindent
To point 2.
\\
As corollary of claim 1, we can rewrite the whole effective Lagrangian expansion as 

\begin{equation}
\mathcal{L}_\text{eff}[h]= M^2 \mathcal{L}^{(-1)}[h, S_0(h)]+\mathcal{L}^{(0)}[h, S_0(h)]+\frac{1}{M^2} \mathcal{L}_\text{eff}^{(1)}[h]+... \, .
\end{equation}
Proving claim 2 is equivalent to proving that $ \mathcal{L}^{(-1)}[h, S_0(h)] = \text{const}$, where $S_0(h)$ is given by (\ref{120}), which is a solution of (\ref{r-1}).
This is not immediately obvious at the first glance, but the decoupling limit tells us that the effective Lagrangian must not contain terms of order $\mathcal{O}(M^2)$.
To prove that $ \mathcal{L}^{(-1)}[h, S_0(h)] = \text{const}$, we must determine  $ \mathcal{L}^{(-1)}[h, S] $ in (\ref{122}).
$\mathcal{O}(M^2)$ terms only enter the Lagrangian through potential terms.
We introduce, in analogy to (\ref{eqn:decomposition}), the decomposition of parameters $\mathcal{O} = M^2 \mathcal{O} ^0+ \bar{\mathcal{O}}$.
We thus have with (\ref{eqn:vevrelation})

\begin{equation}
\begin{aligned}
\mathcal{L}^{(-1)} (h, S)& =  -\left[A^0_{1}\phi_{01}^2 + A^0_2 \phi_{02}^2+B^0 \phi_{01}^2\phi_{02}^2+C^0_1 \phi_{01}^4+ C^0_2 \phi_{02}^4  \right] \\
& = -\Big[B^0(\phi_{01}^2\phi_{02}^2-v_1^2 \phi_{02}^2-v_2^2\phi_{01}^2) +C^0_1 (\phi_{01}^4-2v_1^2\phi_{01}^2) +C^0_2 (\phi_{02}^4-2v_2^2 \phi_{02}^2)\Big]\\
&= B^0v_1^2v_2^2+C^0_1v_1^4+C^0_2 v_2^4  -\Big[B^0(\phi_{01}^2-v_1^2)(\phi_{02}^2-v_2^2)+C^0_1 (\phi_{01}^2-v_1^2)^2 +C^0_2 (\phi_{02}^2-v_2^2)^2\Big]\\
& =\left[B^02v_1^2v_2^2+C^0_1v_1^4+C^0_2 v_2^4\right]-(\phi_{0i}^2-v_i^2)\, \mathcal{D}_{ij}\,(\phi_{0j}^2-v_j^2) \ ,
\end{aligned}
\end{equation}
where we defined 

\begin{equation}
\mathcal{D}=\begin{pmatrix}
C_1^0& B^0/2\\B^0/2 & C^0_2
\end{pmatrix}\,.
\end{equation}
Now, defining a new field $\Xi_i(h, S) = \phi_{0i}^2-v_i^2$, the Lagrangian becomes

\begin{equation}
\begin{aligned}
\mathcal{L}^{(-1)} (h, S)& = -\Xi_i(h,S)\, \mathcal{D}_{ij} \,\Xi_j(h,S)+ \text{const} \\
& = \frac{\Xi_i}{v_i}\, {v_iv_j} \mathcal{D}_{ij}\, \frac{\Xi_j}{v_j}+\text{const}\\
& = -\lim_{M^2\rightarrow \infty } \,\,\frac{1}{8M^2} \bar{\Xi}^T W^{-1T} \mathcal{M}^2W^{-1} \bar{\Xi}+ \text{const} \, ,
\end{aligned}
\end{equation}
where $\bar{\Xi}_i(h,S) = \frac{\Xi_i}{v_i}$ (\ref{eqn:CsB}) has been used.
Defining $\hat{\Xi}(h,S) = W^{-1}\bar{\Xi}$, we obtain 

\begin{equation}
\mathcal{L}^{(-1)}(h, S) = -\lim_{M^2\rightarrow \infty } \,\,\frac{1}{8M^2} \,\hat{\Xi}^T \mathcal{M}^2 \hat{\Xi}+\text{const} = -\frac{1}{8}\,\hat{\Xi_2}^2+ \text{const} \, .
\end{equation}
We note that (\ref{r-1}) comes from the E.O.M. at order $\mathcal{O}(M^2)$, hence it is equivalent to the E.O.M. of $\mathcal{L}^{(-1)}(h,S)$ with respect to $S$:

\begin{equation}
\begin{aligned}
0 &=\frac{\delta \mathcal{L}^{(-1)}}{\delta S}\Bigg|_{S_0} = -\frac{1}{16}\,\hat{\Xi}_2(h,S_0)\, \frac{\partial \hat{\Xi}_2(h,S)}{\partial S}\Bigg|_{S_0} \ .
\end{aligned}
\end{equation}
This equation gives us either

\begin{equation}
\hat{\Xi}_2(h, S_0) =0 \quad \text{or} \quad \partial \hat{\Xi}_2(h,S_0)/\partial S_0 =0 \, .
\end{equation}
The original Lagrangian that depends on $S$ can be written in momentum space as

\begin{equation}
\mathcal{L} \supset \sum_{a}^\infty\sum_i^{N_a} Z_{ai} S^a \, .
\end{equation}
A diagram with $I$ internal lines of $S$, a number $V_{ai}$ of vertex of type $ai$ at tree level satisfies the following topological identities:

\begin{equation}
\begin{aligned}
2I=\sum_a \sum_i a V_{ai} \ , \\
I = \sum_a \sum_i V_{ai}-1 \ .
\end{aligned}
\end{equation}
This gives us 

\begin{equation}
\sum_i^{N_1} V_{1i} = \sum_{a=3} \sum_i (a-2) V_{ai}+2\geq 2 \ .
\end{equation}
Thus, the simplest diagram at tree level with $S$ as internals are formed by combining two vertices of type $Z_{ii}$.
Therefore, the effective Lagrangian resulting from integrating out $S$ at tree level that is supposed to capture this effect locally must start at order $\mathcal{O}(Z_{1i} Z_{1j})$.
To produce this kind of effective action, $S(h)$ must start at order $\mathcal{O}(Z_{1j})$.
That turns $S_0$ to start at $\mathcal{O}(h^2)$.
This consideration helps us to decide that $\hat{\Xi}_2(h, S_0) =0$ is the correct choice, because its solution for $S_0$ starts at $h^2$, while the solution of $\partial \hat{\Xi}_2(h,S_0)/\partial S_0 =0$ for $S_0(h)$ is linear in $h$.
See also the discussion below.
Plugging the value $\hat{\Xi}_2(h,S_0) =0$ back to the Lagrangian $\mathcal{L}^{(-1)}$ confirms claim 2. 

Now, the leading order effective Lagrangian becomes

\begin{equation}
\begin{aligned}
\mathcal{L}_\text{eff}^{(0)} &= \mathcal{L}^{(0)}[h, S_0[h]] = \sum_{a=0}\bar{J}_a S_0^a+\frac{1}{2} \left(1+F_S\right)(\partial S_0)^2+ F_{hS}\,\partial h \,\partial S_0\\
& = \frac{1}{2}(1+F_h(h,S_0)) (\partial h)^2+\frac{1}{2} \left(1+F_S\right)(\partial S_0)^2+ F_{hS}\,\partial h \,\partial S_0 \\
& \quad +\frac{v_1^2}{4}\langle L_\mu L^\mu\rangle \left(1+F_U(h, S_0)\right) + \bar{\psi} M(v+W\bar{\Phi}, \, U) \psi \\
& \quad -(\phi_{0i}^2-v_i^2)\, \bar{\mathcal{D}}_{ij}\,(\phi_{0j}^2-v_j^2)+\left[\bar{B}v_1^2v_2^2+\bar{C}_1v_1^4+\bar{C}_2 v_2^4\right]\ ,
\end{aligned}
\end{equation}
where 

\begin{equation}
\bar{\mathcal{D}} = \begin{pmatrix}
\bar{C}_1&\bar{ B}/2\\\bar{ B}/2 & \bar{ C}_2
\end{pmatrix}\, .
\end{equation}
Note that 

\begin{equation}
\begin{aligned}
\Xi_i\, \bar{\mathcal{D}}_{ij}\,\Xi_j &= \frac{\Xi_i}{v_i}\, v_iv_j\bar{ \mathcal{D}}_{ij} \,\frac{\Xi_j}{v_j} = \frac{1}{8}\lim_{M^2\rightarrow 0} \bar{ \Xi}^T W^{-1T}\mathcal{M}^2W^{-1}\bar{\Xi} =\frac{1}{8}m^2 \hat{\Xi_1}^2 \ .
\end{aligned}
\end{equation}
Therefore, the leading-order effective Lagrangian becomes 

\begin{equation}\label{lo}
\begin{aligned}
\mathcal{L}_\text{eff}^{(0)}& = \frac{1}{2}(1+F_h(h,S_0)) (\partial h)^2+\frac{1}{2} \left(1+F_S\right)(\partial S_0)^2+ F_{hS}\,\partial h \,\partial S_0 \\
& +\frac{v_1^2}{4}\langle L_\mu L^\mu\rangle \left(1+F_U(h, S_0)\right) + \bar{\psi} M(v+W\bar{\Phi}, \, U) \psi -\frac{1}{8}m^2 \hat{\Xi_1}^2(h,S_0)\\
& = \frac{1}{2} (\partial \tilde{h})^2+\frac{v_1^2}{4}\langle L_\mu L^\mu\rangle \left(1+F_U(h, S_0)\right)  -\frac{1}{8}m^2 \hat{\Xi_1}^2(h,S_0)+ \bar{\psi} M(v+W\bar{\Phi}, \, U) \psi\, ,
\end{aligned}
\end{equation}
where we have absorbed the derivative interactions by a field redefinition

\begin{equation}\label{eq:htilde}
\frac{1}{2} \left(1+F_h\right)(\partial h)^2+\frac{1}{2} \left(1+F_S\right)(\partial S_0)^2+  F_{hS}\,\partial h \,\partial S_0 \longrightarrow \frac{1}{2} (\partial \tilde{h})^2 \ .
\end{equation}
This field redefinition modifies $F_U$, $V(h)$ and the Yukawas.

The expression for the field $\hat{\Xi}_i(h, S)$ was computed using Wolfram Mathematica \cite{Mathematica} and the result is (keep in mind the shorthands \eqref{eqn:shorthands}) 

\begin{equation}
\begin{aligned}
\hat{\Xi}_1(h, S)& = \frac{1}{v_1v_0}\Bigg\{\left(\frac{fv_1\cos(\theta)}{u}-v_2 \sin(\theta)\right)
\left(\frac{\left(N_1(h, S)\right)^2}{f^2v_0^2}-v_1^2\right)
\\
& \quad +\frac{v_1}{v_2}\left(\frac{fv_2\cos(\theta)}{u}+v_1 \sin(\theta)\right)\left(\frac{\left(N_2(h, S)\right)^2}{f^2v_0^2}-v_2^2\right)\Bigg\}\\
& \doteq G_1 N_1^2+G_2 N_2^2+L \ , 
\end{aligned}
\end{equation}
\begin{equation*}
\begin{aligned}
\hat{\Xi}_2(h, S)& = \frac{1}{v_1v_0}\Bigg\{\left(\frac{fv_1\sin(\theta)}{u}+v_2 \cos(\theta)\right)
\left(\frac{\left(N_1(h, S)\right)^2}{f^2v_0^2}-v_1^2\right)\\
& \quad +\frac{v_1}{v_2}\left(\frac{fv_2\sin(\theta)}{u}-v_1 \cos(\theta)\right)\left(\frac{\left(N_2(h, S)\right)^2}{f^2v_0^2}-v_2^2\right)\Bigg\}\\
&\doteq D_1 N_1^2+D_2 N_2^2+E \ ,
\end{aligned}
\end{equation*}
where 

\begin{align}
N_1(h,S)&=h\Big[v_1u\,\cos(\theta)-fv_2\sin(\theta)\Big]+S\Big[v_1u\,\sin(\theta)+fv_2\cos(\theta)\Big]+fv_1v_0 \nonumber \\
&\doteq \alpha_1 h+\beta_1 S+\gamma_1 \ , \nonumber \\
N_2(h,S) &= h\Big[v_2u\,\cos(\theta)+fv_1\sin(\theta)\Big]+S\Big[v_2u\,\sin(\theta)-fv_1\cos(\theta)\Big]+fv_2v_0 \nonumber \\
&\doteq \alpha_2 h+\beta_2 S+\gamma_2 \ , \nonumber \\
G_{1} & = \frac{1}{f^2v_1v_0^3}\left(\frac{fv_1\cos(\theta)}{u}-v_2 \sin(\theta)\right) \ , \nonumber \\
G_{2} &=  \frac{1}{f^2v_2v_0^3} \left(\frac{fv_2\cos(\theta)}{u}+v_1 \sin(\theta)\right) \ , \nonumber \\
L &=  -\frac{v_1}{v_0}\left(\frac{fv_1\cos(\theta)}{u}-v_2 \sin(\theta)\right)  - \frac{v_2}{v_0}  \left(\frac{fv_2\cos(\theta)}{u}+v_1 \sin(\theta)\right) \ , \nonumber \\
D_{1} & = \frac{1}{f^2v_1v_0^3}\left(\frac{fv_1\sin(\theta)}{u}+v_2 \cos(\theta)\right) \ , \nonumber \\
D_{2} &=  \frac{1}{f^2v_2v_0^3}  \left(\frac{fv_2\sin(\theta)}{u}-v_1 \cos(\theta)\right) \ , \nonumber \\
E &=  -\frac{v_1}{v_0}\left(\frac{fv_1\sin(\theta)}{u}+v_2 \cos(\theta)\right)  - \frac{v_2}{v_0}  \left(\frac{fv_2\sin(\theta)}{u}-v_1 \cos(\theta)\right) \ .
\end{align}
We obtain the closed-form solution for $S_0(h)$ by solving the quadratic equation $\hat{\Xi}_2(h,S_0)=0$, giving (the other solution with positive sign +1 does not start at order $\mathcal{O}(h^2)$, so we reject it)

\begin{equation}\label{S00}
\begin{aligned}
S_0(h)&=\frac{b}{2a}\Bigg(\left(1-\frac{4ac}{b^2}\right)^{1/2}-1 \Bigg) = \sum_{k=2} r_k h^k \ , 
\end{aligned}
\end{equation}
where

\begin{equation}
\begin{aligned}
a &=D_1 \beta_1^2+D_2 \beta_2^2 \ , \\
b& = 2\left(D_1 \alpha_1 \beta_1+D_2 \alpha_2\beta_2\right)h+{\left[2D_1\beta_1\gamma_1+2D_2\beta_2\gamma_2\right]} = 2\left(D_1 \alpha_1 \beta_1+D_2 \alpha_2\beta_2\right)h+1 \ , \\
c& =D_1\left(\alpha_1 h+\gamma_1\right)^2+D_2\left(\alpha_2 h+\gamma_2\right)^2+E = \left(D_1\alpha_1^2+D_2 \alpha_2^2\right) h^2 \ .
\end{aligned}
\end{equation}
Note that we have used the relations

\begin{equation}
2D_1\beta_1\gamma_1+2D_2\beta_2\gamma_2 = 1 \ , \quad D_1\alpha_1 \gamma_1+D_2\alpha_2\gamma_2=0 \quad \text{and} \quad D_1\gamma_1^2+D_2\gamma_2^2+E=0 \, ,
\end{equation}
which can be verified easily by direct computation.
One can expand the closed-form solution (\ref{S00}) and check whether it starts at $\mathcal{O}(h^2)$,

\begin{equation}
S_0(h) = -\frac{c}{b}+\frac{ac^2}{b^3}-\frac{2a^2c^3}{b^5}+...=-(D_1\alpha_1^2+D_2 \alpha_2^2)\,h^2+... \, .
\end{equation}
The other coefficients of $h^k$ can be computed from the expansion.
The results are 

\begin{equation}
\begin{aligned}
r_2 &= -(D_1\alpha_1^2+D_2 \alpha_2^2) \ , \\
r_3 & ={2\left(D_1 \alpha_1 \beta_1+D_2 \alpha_2\beta_2\right) \left(D_1\alpha_1^2+D_2 \alpha_2^2\right) } \ , \\
r_4 & ={\left(D_1 \beta_1^2+D_2 \beta_2^2\right)\left(D_1\alpha_1^2+D_2 \alpha_2^2\right)^2}  -{\left(D_1\alpha_1^2+D_2 \alpha_2^2\right) 4\left(D_1 \alpha_1 \beta_1+D_2 \alpha_2\beta_2\right)^2} \ , \\
r_5 & ={\left(D_1\alpha_1^2+D_2 \alpha_2^2\right) 8\left(D_1 \alpha_1 \beta_1+D_2 \alpha_2\beta_2\right)^3 }   -{3\left(D_1 \beta_1^2+D_2 \beta_2^2\right)2\left(D_1 \alpha_1 \beta_1+D_2 \alpha_2\beta_2\right) \left(D_1\alpha_1^2+D_2 \alpha_2^2\right)^2} . \\
\end{aligned}
\end{equation}
Having found the explicit form of $S_0(h)$, one can immediately perform the field redefinition to absorb the derivative interactions.
Specifically, we have with \eqref{eq:htilde}

\begin{equation}
\begin{aligned}
& \frac{1}{2} \left(1+F_h\right)(\partial h)^2+\frac{1}{2} \left(1+F_S\right)(\partial S_0)^2+  F_{hS}\,\partial h \,\partial S_0  \doteq \frac{1}{2}\left(1+\right)(\partial h)^2 \doteq \frac{1}{2} (\partial \tilde{h})^2 \ ,
\end{aligned}
\end{equation}
with a new function $F(h)$ and a new field $\tilde{h}$ that can be expressed as a polynomial expansion in $h$, 

\begin{equation}
\tilde{h}(h)=\int^h_0\sqrt{1+F(s)}\,\,ds = h\left(1+ \sum_{k=1} w_k h^k\right) \, .
\end{equation}
Using Wolfram Mathematica \cite{Mathematica}, we find the first few of the coefficients $w_k$:

\begin{equation}
\begin{aligned}
w_1 &=  \frac{ v_0 \,\cos\theta}{2 f u} \ , \\
w_2 &=\frac{2uv_0^2\,\cos^2\theta+f^2u(1+4r_2^2u^2)+4fr_2u^2v_0\sin\theta}{6f^2u^3} \ , \\
w_3&=\frac{1}{16f^3u^3}\Bigg[24f^3r_2r_3u^3+3\cos \theta\left(v_0(2f^2+v_0^2)+4fr_2u v_0^2\sin\theta\right) \\
& \quad +12f^4v_0 r_3\sin\theta+v_0^3\cos(3\theta) -12v_0^3f^2r_3 \sin\theta \Bigg] \, .
\end{aligned}
\end{equation}
This equation can be inverted to obtain $h$ as a function of $\tilde{h}$:

\begin{equation}\label{12345}
h (\tilde{h}) = \sum_{k=1} z_k \tilde{h}^k \ .
\end{equation}
The first few coefficients are computed to

\begin{equation}
\begin{aligned}
z_1&=1 \ , \\
z_2 &=-w_1=- \frac{ v_0 \,\cos\theta}{2 f u} \ , \\
z_3 &= 2w_1^2-w_2=\frac{uv_0^2\,\cos^2\theta-f^2u(1+4r_2^2u^2)-4fr_2u^2v_0\sin\theta}{6f^2u^3} \ .
\end{aligned}
\end{equation}
Now, let us write the ultimate leading order effective action as 

\begin{equation}
\mathcal{L}_\text{eff}^{(0)}(\tilde{h})=\frac{1}{2} (\partial \tilde{h})^2+\frac{v_1^2}{4}\langle L_\mu L^\mu\rangle \left(1+\tilde{F}_U(\tilde{h})\right) -\tilde{V}(\tilde{h})+  \bar{\psi} \tilde{M}(\tilde{h}, \, U) \psi\, .
\end{equation}
We write

\begin{equation}
F_U(h,S) = F_1 h+F_2 S+F_{11} h^2+ F_{12} hS+F_{22} S^2 \ .
\end{equation}
Then, after integrating out $S$ and inserting $S= S_0$ we get 

\begin{equation}
\begin{aligned}
F_U(h(\tilde{h})) =& F_1 h+\left(F_{11}+r_2 F_2\right)h^2+ (F_{12} r_2 + F_2 r_3)h^3 +(F_{22} r_2^2 + F_{12 }r_3 + F_2 r_4)h^4 +... \, .
\end{aligned}
\end{equation}
Replacing $h$ by (\ref{12345}), one finds

\begin{equation}
\begin{aligned}
\tilde{F}_U(\tilde{h}) &= F_1 \tilde{h} +  \Big[F_{11} + F_2 r_2 + F_1 z_2\Big]\tilde{h}^2 + \Big[F_{12} r_2 + F_2 r_3 + 2 (F_{11} + F_2 r_2) z_2 + F_1 z_3\Big]\tilde{h}^3 +\mathcal{O}(\tilde{h}^4)\\
&= \tilde{F}_1 \tilde{h}+ \tilde{F}_2 \tilde{h}^2+\tilde{F}_3 \tilde{h}^3+... \, .
\end{aligned}
\end{equation}
We compute

\begin{equation}\label{eqn:Fs}
\begin{aligned}
\tilde{F}_1 &=\frac{2(uv_1\cos\theta-fv_2\sin\theta)}{fv_0v_1} \ , \\
\tilde{F}_2  &=\frac{uv_1\cos\theta-fv_2\sin\theta}{f^2v_1^2}
\\
& \times \Bigg[-\frac{v_1\cos\theta}{u}+\frac{2v_0^2\sin\theta\,\left(fv_1\sin\theta+uv_2\cos\theta\right)\left(uv_1\sin\theta+fv_2\cos\theta\right)}{uv_2v_0^4}+\frac{uv_1\cos\theta-fv_2\sin\theta}{v_0^2}\Bigg]  , 
\end{aligned}
\end{equation}
\begin{equation*}
\begin{aligned}
\tilde{F}_3 &= \frac{u v_1 \cos\theta-f v_2 \sin\theta}{3 f^3 u^2 v_0^7 v_1^3 v_2^2} 
\\
& \times \Big(6 u v_0^2 v_2 v_0^2 \sin\theta (u v_2 \cos\theta+f v_1 \sin\theta) (f v_2 \cos\theta+u v_1 \sin\theta) (u v_1 \cos\theta-f v_2 \sin\theta)\\
&-3 v_0^4 v_1 v_2 \cos\theta (2 v_0^2 \sin\theta (u v_2 \cos\theta+f v_1 \sin\theta) (f v_2 \cos\theta+u v_1 \sin\theta)+u v_0^2 v_2 (u v_1 \cos\theta-f v_2 \sin\theta))\\
&-6 v_0^4 (u v_2 \cos\theta+f v_1 \sin\theta)^2 (f v_2 \cos\theta+u v_1 \sin\theta) (u v_1 \cos\theta-f v_2 \sin\theta) \sin(2 \theta)\\
&-u^2 (f^2 v_0^6 v_1^2 v_2^2-v_0^8 v_1^2 v_2^2 \cos\theta^2
\\
& +2 v_0^4 v_1 v_2 \sin\theta^2 (v_1 v_2 v_0^2 (-f^2+u^2+(f^2+u^2) \cos(2 \theta))+f u (v_1^4-v_2^4) \sin(2 \theta))\\
&+\sin\theta^2 (v_1 v_2 v_0^2 (-f^2+u^2+(f^2+u^2) \cos(2 \theta))+f u (v_1^4-v_2^4) \sin(2 \theta))^2) \Big) \ .
\end{aligned}
\end{equation*}

The potential terms can be read off from (\ref{lo}) and the subsequent equations.
They take the form

\begin{align}
& V^{(0)}(h, S)  = \frac{1}{8}m^2\hat{\Xi}_1^2(h, S)\nonumber\\
& = \frac{1}{8}m^2\Bigg\{\Big[G_1 \beta_1^2+G_2 \beta_2^2\Big] S^2+ 2\Big[G_1 \alpha_1 \beta_1+G_2 \alpha_2\beta_2\Big]hS  +2 h+ \Big[G_1\alpha_1^2+G_2 \alpha_2^2\Big] h^2\Bigg\}^2\nonumber\\
& = \frac{1}{8}m^2\Bigg\{ \Big[G_1 \beta_1^2+G_2 \beta_2^2\Big]^2\, S^4+ \Big[G_1\alpha_1^2+G_2 \alpha_2^2\Big]^2\, h^4\nonumber\\
&\quad\quad+\Big(4 \Big[G_1 \alpha_1 \beta_1+G_2 \alpha_2\beta_2\Big]^2+2  \Big[G_1 \beta_1^2+G_2 \beta_2^2\Big]\Big[G_1\alpha_1^2+G_2 \alpha_2^2\Big]\Big)h^2S^2
\nonumber\\
&\quad\quad+4\Big[G_1 \beta_1^2+G_2 \beta_2^2\Big]\Big[G_1 \alpha_1 \beta_1+G_2 \alpha_2\beta_2\Big]hS^3  + 4\Big[G_1 \alpha_1 \beta_1+G_2 \alpha_2\beta_2\Big]\Big[G_1\alpha_1^2+G_2 \alpha_2^2\Big] h^3S\nonumber\\
& \quad\quad+ 4\Big[G_1 \beta_1^2+G_2 \beta_2^2\Big]h S^2+ 8\Big[G_1 \alpha_1 \beta_1+G_2 \alpha_2\beta_2\Big]h^2S +4\Big[G_1\alpha_1^2+G_2 \alpha_2^2\Big] h^3+4h^2\Bigg\}\nonumber\\
& \doteq \frac{1}{8}m^2\Big(P_s S^4+P_h h^4+Q_{hhss}h^2S^2+Q_{hsss}hS^3+Q_{hhhs}h^3S  +T_{hss} hS^2+T_{hhs}h^2S+T_h h^3+4h^2\Big) \, ,
\end{align}
where we have used the fact that

\begin{equation}
    G_1\alpha_1\gamma_1+G_2\alpha_2\gamma_2=1 \ , \quad G_1 \beta_1\gamma_1+G_2\beta_2\gamma_2 =0 \quad \text{and} \quad G_1\gamma_1^2+G_2\gamma_2^2+L=0 \, .
\end{equation}
Note that we obtain the form of the potential that we expect.
No linear terms in $h$ or $S$ and no mass mixing terms $hS$ appear in the potential.
Note also that we do not have a mass term $M^2S^2$ here, because this term belongs to $\mathcal{L}^{(-1)}_\text{eff}$, which together with other terms have been shown to vanish in claim 2.
Next, we insert the classical solution $S_0(h)$ into the potential:

\begin{equation}
\begin{aligned}
&V^{(0)}(h, S_0(h)) =\frac{1}{8}m^2 \Bigg[4h^2+T_h h^3 + (P_h + r_2 T_{hhs}) h^4 +(r_2Q_{hhhs}  + r_3 T_{hhs} + r_2^2 T_{hss}) h^5 +\mathcal{O}(h^6)
\Bigg] .
\end{aligned}
\end{equation}
Finally, we re-express it in terms of the new field $\tilde{h}$.
The result is 

\begin{equation}
\begin{aligned}
 \tilde{V}(\tilde{h}) & =\frac{1}{8}m^2 \Bigg[ 4 \tilde{h}^2 + (T_h + 8 z_2) \tilde{h}^3 + (P_h + r_2 T_{hhs} + 3 T_h z_2 + 4 z_2^2 + 8 z_3) \tilde{h}^4 +\mathcal{O}(\tilde{h}^5)\Bigg]\\
& = \frac{1}{2}m^2\tilde{h}^2+V_3 \,\tilde{h}^3+ V_4\,\tilde{h}^4+... \, .
\end{aligned}
\end{equation}
We compute

\begin{equation}\label{eqn:Vs}
\begin{aligned}
V_3 & =\frac{m^2}{8fuv_0^3}\left[-4v_0^4\cos\theta+\frac{v_0^2}{v_1v_2}\left((f^2+5u^2)v_1v_2\cos\theta-(f^2+u^2)v_1v_2\cos(3\theta)+4uf(v_1^2-v_2^2)\sin^3(\theta)\right)
\right]  ,
\end{aligned}
\end{equation}
\begin{equation*}
\begin{aligned}
V_4 & =
\frac{m^2}{48 f^2 u^2 v_0^6 v_1^2 v_2^2}
\\
& \times \Bigg(6 v_0^8 v_1^2 v_2^2 \cos(\theta)^2-48 (v_1^2+v_2^2)^2 \sin(\theta)^2 (u v_2 \cos(\theta)+f v_1 \sin(\theta)) (f v_2 \cos(\theta)+u v_1 \sin(\theta)) (u v_1 \cos(\theta)\\
&-f v_2 \sin(\theta)) (f v_1 \cos(\theta)-u v_2 \sin(\theta))+6 (v_1^2+v_2^2)^2 (u^2 v_1 v_2 \cos(\theta)^3+(f^2+2 u^2) v_1 v_2 \cos(\theta) \sin(\theta)^2\\
&+f u (v_1^2-v_2^2)\sin(\theta)^3)^2-9 v_0^4 v_1 v_2 (v_1^2+v_2^2) \cos(\theta) ((f^2+5 u^2) v_1 v_2 \cos(\theta)-(f^2+u^2) v_1 v_2 \cos(3\theta)\\
&+4 f u (v_1-v_2) (v_1+v_2) \sin(\theta)^3)-8 (f^2 v_0^6 v_1^2 v_2^2-v_0^8 v_1^2 v_2^2 \cos(\theta)^2\\
&+2 v_0^4 v_1 v_2 \sin(\theta)^2 (v_1 v_2 (v_1^2+v_2^2) (-f^2+u^2+(f^2+u^2) \cos(2\theta))+f u (v_1^4-v_2^4) \sin(2\theta))\\
&+\sin(\theta)^2 (v_1 v_2 (v_1^2+v_2^2) (-f^2+u^2+(f^2+u^2) \cos(2\theta))+f u (v_1^4-v_2^4) \sin(2\theta))^2)\Bigg) \ .
\end{aligned}
\end{equation*}
We note here that the vev of the field $ h $ after integrating out $ S $ is not $ v_1 = v= 246 \, GeV $.
Even before integrating out $S$ the physical field $ h $ does not describe the excitation around the vacuum at $ v_1 $ and vice versa for $ S $ and $ v_2 $, see Equation (\ref{eqn:originalphysical}).
Instead, we are dealing with mixed states.
We cannot recover a very SM-like situation unless we let the second scalar decouple fully.

Let us have a look at viable values of the parameter space.
We want to approach the Standard Model.
First, we try a vanishing mixing angle.
Setting $ \sin \theta =0 $, implying \eqref{eqn:Mbar} to be already diagonal, gives

\begin{equation}
\begin{aligned}
\tilde{F}_1 = 2 \frac{u}{f v_0} = 2 \sqrt{\frac{1}{v_0^2} - \frac{1}{f^2}} \ , \\
\tilde{F}_2 = -\frac{1}{f^2} + \frac{u^2}{f^2 v_0^2} = \frac{1}{v_0^2} - \frac{2}{f^2} \ .
\end{aligned}
\end{equation}
We arrive at the most "SM-like" behavior if also $ v_2 \rightarrow 0 $.
We then replace $ v_0 \rightarrow v_1= v $.
Compare this to the SM values $ F_1 = 2/v \, , \ F_2 = 1/v^2 $.
The anomalous effect is thus of order of the inverse of the composite Higgs scale $ f \ (f^2) $.
The discussed limit is problematic, though, since for $ v_2 \rightarrow 0 $ we have blowing up potential parameters \eqref{eqn:Vs} and interaction functions $ \tilde{F} $ \eqref{eqn:Fs} as a result of \eqref{eqn:vevrelation}, where we have made the point about $ v_2 $.
It is still possible to let the model \textit{approach} the Standard Model (with non-zero $ v_2 $) whereas before integrating out the field $ S $ there was no such close connection of $ h $ to the SM Higgs.

Even though there is a bunch of tunable parameters, any other limit soon contradicts observations.
This can be demonstrated at the hand of $ \tilde{F}_1 $.
It can be written as

\begin{equation}
\tilde{F}_1 = \frac{2}{v_1} \dfrac{uv_1 \cos \theta - f v_2 \sin \theta}{f v_0} \ .
\end{equation}
In the SM limit the right fraction becomes equal to one.
However,

\begin{equation}
\dfrac{uv_1 \cos \theta - f v_2 \sin \theta}{f v_0} = \sqrt{1-\frac{v_0^2}{f^2}} \frac{v_1}{v_0} \cos \theta - \frac{v_2}{v_0} \sin \theta < 1 \ .
\end{equation}

The SM limit of the $ SO(6)/SO(5) $ CHM with the Lagrangian \eqref{eqn:electroweakSO6} is obtained by letting the other scalar $ \eta_0 $ decouple.
To avoid mixings of the scalars, only \textit{the} Higgs must obtain a vev, the potential parameter $ B $ has to be put equal to zero and $ f $ must be large enough for a sufficient suppression of the derivative interactions of the two scalars.
In this limit, the physical fields are easily identified.
The Higgs potential $ V = A_1 \xi_0^2 + C_1 \xi_0^4 $ permits $v= \sqrt{-A_1/2C_1} $ which sets the electroweak scale in \eqref{eqn:electroweakSO6}.
We then identify the observed SM Higgs as $ h = (\xi_0-v)/\sqrt{1-v/f} $ (canonically normalized) with the mass $ m_h^2= -4 A_1 $.
This fixes the Higgs potential parameters.
The decoupling scalar $ S= \eta_0 $ has the mass $ m_S^2= 2 A_2 $ which is unconstrained just like the parameter of the quartic $ S $ term $ C_2 $.

\section{Outlook -- Expectations and Opportunities}\label{sec:outlook}
The SM contains only a complex phase in the CKM matrix as source for CP violation.
Even if there was strong CP violation and on top CP violation in the neutrino sector beyond the SM, it would almost certainly not sum up to the amount needed to realize the observed matter- antimatter asymmetry.
Moreover, for electroweak baryogenesis the phase transition in the early universe by electroweak symmetry breaking had to be strongly first-order.
In other words, fast enough to prevent from (sphaleron \cite{sphaleron}) effects which compensate the asymmetry.
Here, too, the SM fails \cite{SMEWPht}.
We argued in the motivation for an additional scalar that it might serve as source for more CP violation.
Indeed it is found that a SM-gauge singlet scalar can provide all necessities for a strong first-order phase transition leading to a viable baryogenesis \cite{EWPhTsinglet}.
The most interesting scenario occurs when both singlet fields of the $ SO(6)/SO(5) $ CHM obtain vevs and the Higgs states mix.
We then not only have the typical non-linear composite Higgs corrections to couplings, but also CP violation in the Higgs sector.
CP is then spontaneously broken by the Higgs potential, too.
Specifically, we have a dimension five operator describing a coupling of the Higgs and the pseudo-scalar singlet to a pair of tops describing the interaction of Equation \eqref{eqn:yuks}.
It is a dimension five operator that in Standard Model effective field theory takes the form \cite{EWPhT}

\begin{equation}
\frac{c}{f} \, is \, H \bar{Q}_3 \gamma^5 t \ +\text{h.c.}
\end{equation}
$ \bar{Q}_3 $ denotes the third quark generation doublet.
It does not explicitly violate CP.

Notably, the additional singlet is a simple and effective candidate to structure the scalar (Higgs) potential in such a way that the electroweak phase transition is indeed strong, $ v/T_c >1 $.
This is because the new member of the Higgs sector can contribute to the potential directly via tree level instead of the usual thermal loop effects that only lead to critical values of the order parameter \cite{EWPhTsinglet,SMEWPht}.

All in all, the model contains the demanded characteristics for baryogenesis for natural values of its parameters.
There are, however, subtleties concerning the vev of the additional scalar $ s $, which is demanded to vary during the phase transition and maybe even to have vanished to its end, and the need for a small source of explicit CP violation in the scalar sector \cite{EWPhT}.

There might still be explicit CP violation present in the strong sector itself, in the couplings of the Goldstones to the fermionic partners \cite{Redi}.
Yukawa couplings that also vary during the phase transition can serve as a source and have been studied in a review on electroweak phase transition in composite Higgs models \cite{EWPhTbaryocomposite}.

As for baryogenesis, there have been attempts to solve likewise the dark matter problem by adding a gauge-singlet scalar to the SM \cite{minimaldm,colddm,singletdark}.
Again, there is the $ SO(6)/SO(5) $ CHM at hand.
To make $ s $ stable, we can demand a parity transformation $ s \rightarrow -s $ which must be a symmetry of the full strong sector \cite{compositedark}.
If there is such a $ Z_2 $ symmetry present enlarging the coset to $ O(6)/ O(5) $
with the six-dimensional parity acting as

\begin{equation}
\hat{P}_6 = (1,1,1,1,-1,1)^T \ ,
\end{equation}
$ s $ could be a dark matter candidate.
In fact, we pointed out in Section \ref{so6goldstone} that we have this symmetry for our choice of embedding.
The vevs of the scalars might spoil the structure of the potential.
Then, the symmetry has to be enforced by hand, but potential and embedding is up to speculation anyhow, so we concentrate on the direct implications of the model's coset for this outlook section.
Since the odd $ Z_2 $ parity is required for its stability as a the dark matter particle, $ s $ is not supposed to obtain a vev that would break the symmetry spontaneously \cite{dm,compositedark}.
There would thus be no mixing of Higgs states.
Without mixing, it follows that there is no linear coupling to SM gauge bosons.
Furthermore, we have to demand $ \epsilon = 0 $, i.e. $ \hat{P}_6 \, q_R = +1 \, q_R $, such that there is no mixing in the 5-6 components of the right-handed fermions that violate the $ Z_2 $ symmetry in the couplings to $ s $ \cite{Redi}.
Hence, there is no linear coupling to SM fermions, see the end of Section \ref{fermion}.
Since it would still be massive, the criteria to be a dark matter candidate are fulfilled.

The nonrenormalizable structures in the CHM -- derivative interactions of $ s $ with the Higgs that are determined by the coset structure and the non-linear couplings of $ s $ to the SM fermions -- can serve as discriminators to other models with a dark matter singlet scalar \cite{dm}.

It is found in scans \cite{Redi} and the calculation of the Coleman-Weinberg potential \cite{dm} that the pseudo-scalar singlet is naturally heavier than the Higgs.
Whereas the Higgs vev is fixed and has to be fine-tuned at least to some degree in the CHM, the other vev is free to follow naturalness.
A natural mass for a Goldstone of strong dynamics with a scale $ f\lesssim 1TeV $ could be about $ \sim 500 GeV $.
Of course, when $ \epsilon \rightarrow  0$, $ s $ becomes lighter and eventually massless.
Technically, $ \epsilon \rightarrow 0 $ is technically unnatural again.

The quadratic divergence to the Higgs mass generated by the top loops must be cut-off around (below) the TeV scale to account for a small tuning of $ \sim 10\% $.
Thus the discovered light Higgs requires in the composite Higgs framework fermionic partners (resonances) necessarily lighter than one TeV to generate the correct Higgs mass without too much fine-tuning \cite{lightpartners,Redi}.
This gives the LHC the opportunity to detect or exclude them (exclusion at least in some mass range) in the near future.
The vector resonances can still be assumed to appear at $ \Lambda = 4\pi v \approx 3 TeV $.
Hence, this is one of the models where new physics can be expected at the TeV level.

Since evidence of new physics in the form of directly produced and observed heavier degrees of freedom is missing, it is mandatory to concentrate for the time being on indirect hints to effects of virtual new physics particles.
The measurement of FCNC processes in this context is very appealing since in the Standard Model they are strongly suppressed by the GIM mechanism \cite{gim}.
The Higgs sector can affect those, e.g., via loops.
We can read off from our results that there are four-fermion divergences.
If the parameters are not such that these divergences vanish, we have accordingly FCNC next-to-leading order contributions.
In the case of one Higgs, there are no such contributions in the limit of SM parameters \cite{guo}.
But the $ SO(6)/SO(5) $ CHM inhabits an immediate parameter that could trigger FCNC processes at tree level;
if the mixing parameter $ \epsilon $ is not family symmetric, the singlet $ s $ can mediate FCNC processes.
It can then also decay in a family violating way \cite{gripaios}.
However, if it is of substantial mass, it decays into a top quark pair with a branching ratio of Br $ \sim 100  \% $ \cite{chala}.

Ever since the discovery of the "Higgs-like" particle, there is strong focus on more accurate measurements of its properties -- a discrimination between a SM Higgs and a composite one is possible \cite{chsearch, fingerprinting, carmi, SILHorig}.

\ \\
\textbf{Note}

The present work is based on our master's theses \cite{faiq,andi}.
A very similar approach of matching the next-to-minimal composite Higgs model (CHM) to the electroweak chrial theory has been taken in \cite{qi2021effective}.
We see our work as complementary to it as we use previously developed techniques to extract divergences of the Lagrangian.
Therefore, our construction is implemented such that it fits to our generalized renormalization procedure derived in \cite{lindnermuzakka2022}.
The methods there are derived in order renormalize any pure scalar extension to the electroweak chiral theory to one loop in conjunction with the results for the non-scalar sector calculated in \cite{completerenorm}.
This incorporates any CHM, so the next-to-minimal CHM is exemplary in this work.
The renormalization of the miminal model with coset structure SO(5)/SO(4) has indirectly been carried out with the matching and renormalization of the electroweak chiral Lagrangian with a Higgs \cite{Krause,completerenorm,guo,rgehiggseft}.

As the procedure depends on a polynomial expansion of the scalar fields, this renders our terms more cumbersome overall.

\section*{Acknowledgments}
We want to thank Gerhard Buchalla for the supervision of our Master's theses that this work is based on.
We are also grateful for the discussions with him during the production process of this paper.

\section*{Supplementary Material}
The Mathematica notebook at \href{https://notebookarchive.org/2022-04-1fbhb4o}{https://notebookarchive.org/2022-04-1fbhb4o} is a collection of some supplementary analyses for the joint project of this work and the renormalization formula derived in \cite{lindnermuzakka2022}.
There are chapters with assisting and extra calculations for the spurion analysis, the diagonalization and canonicalization of the model, and divergence extraction.

\appendix

\section{Group Theory}
\label{grouptheory}

\subsection{SO(4) and SU(2) }
\label{SO4SU2}

A few words on the isomorphism $ SO(4) \simeq SU(2)_L \times SU(2)_R $.
We can associate a real vector in the fundamental \textbf{4} representation of SO(4)
$ h^a = (h^1,h^2,h^3,h^4)^T $ with a $ 2 \times 2 $ pseudo-real\footnote{$  \Phi^* = \sigma^2 \Phi \sigma^2 $.} matrix

\begin{equation}
\Phi = \frac{1}{\sqrt{2}} \ ( i \sigma^i h^i + \mathbf{1}_2 h^4 ) \doteq \frac{1}{\sqrt{2}} \ \lambda^a h^a \, ,
\end{equation} 
with the three Pauli-matrices $ \sigma^i $ and the definition $ \lambda^a = (i \sigma^a, \mathbf{1}_2 )$.

The group SO(4) is a rotation on the vector $ h^a $ that preserves its norm,

\begin{equation}
h^a \rightarrow R^{ab} h^b \ , \quad R \in SO(4) \ .
\end{equation}
On the matrix $ \Phi $ the chiral group $ SU(2)_L \times SU(2)_R $ acts as left- and right-multiplication, such that the determinant of $ \Phi $ is unchanged,

\begin{equation}
\Phi \rightarrow g_L \Phi g_R^\dagger \ , \quad g_L \in SU(2)_L \ , \ g_R \in SU(2)_R \ .
\label{eqn:chiraltrafo}
\end{equation}
$ \Phi $ is in the (\textbf{2},\textbf{2}) representation (a pseudo-real bi-doublet) of the chiral group.

Since

\begin{equation}
|h^a|^2 = \langle \Phi^\dagger \Phi \rangle \ ,
\end{equation}
and the trace is invariant under the transformation (\ref{eqn:chiraltrafo}), we find that the chiral transformations leave the norm of $ h^a $ unchanged.
$ SO(4) $ contains the most general norm-preserving transformation of the four-vector.
This shows that a chiral transformation is an element of $ SO(4) $.
Therefore, also the chiral group algebra is contained in the $ SO(4) $ one.
Both have dimension 6.
No sub-algebra (other than the full algebra) exists, with the same dimensionality \cite{Wulzer}.
We thus identify \textbf{4} = (\textbf{2},\textbf{2}).

For each $ SO(4) $ rotation $ R $ on $ h^a $ there are two $ SU(2)_L \times SU(2)_R $ rotations acting in the same way on $ \Phi $: $ g_L,g_R $ and $ -g_L, -g_R $.
We can thus identify \cite{contino}

\begin{equation}
SO(4) = \dfrac{SU(2)_L \times SU(2)_R}{Z_2} \ .
\end{equation}
We now turn to the explicit construction of the $ SO(4) $ generators

\begin{equation}
T^a_{ij} =(T_{L,ij}^a,T_{R,ij}^a) \, , \ a=1,2,3 \ ,
\end{equation}
such that

\begin{equation}
[T_L^a,T_L^b] = i \epsilon^{abc} T_L^c \ , \ [T_R^a,T_R^b] = i \epsilon^{abc} T_R^c \ , \ [T_L^a,T_R^b] = 0 \ ,
\end{equation}
in the correct $ SU(2)_L \times SU(2)_R $ manner.

Consider chiral transformations $ g_{L,R} = 1 + i \delta_{L,R}^\alpha \frac{\sigma^\alpha}{2} +... $ on $ \Phi $.
The infinitesimal variations are

\begin{equation}
\delta_L \Phi = i \delta_L^\alpha \frac{\sigma^\alpha}{2} \Phi \quad \textrm{and} \quad \delta_R \Phi = - i \delta_R^\alpha \frac{\sigma^\alpha}{2} \Phi \ .
\end{equation}
The corresponding variations of $ h^a $ are,

\begin{equation}
\delta_L h^a = i \delta_L^\alpha T^\alpha_{L,ab} \ h^b \quad \textrm{and} \quad  \delta_R h^a = i \delta_R^\alpha T^\alpha_{R,ab} \ h^b \ .
\end{equation}
Combining this yields, for the case of a left transformation,

\begin{equation}
\delta_L \Phi = i \delta_L^\alpha \frac{\sigma^\alpha}{2} \frac{1}{\sqrt{2}} \lambda_b h^b = i \delta_L^\alpha \frac{1}{\sqrt{2}} \lambda_i T^\alpha_{L,ij} h^j \ .
\label{eqn:varyleft}
\end{equation}
The $ \lambda^a $ fulfill the normalization condition $ \langle \lambda_a^\dagger \lambda_b \rangle = 2 \delta_{ab} $.
We make use of this by multiplication with $ \lambda_a^\dagger $ from the left on (\ref{eqn:varyleft}) and then taking the trace.
We obtain

\begin{equation}
i \delta_L^\alpha \frac{1}{\sqrt{2}} \ \frac{1}{2} \langle \lambda_a^\dagger \sigma^\alpha \lambda_b \rangle h^b   = i \delta_L^\alpha \frac{1}{\sqrt{2}} \ 2 \delta_{ai} T_{L,ij}^\alpha h^j \ ,
\end{equation}
from which follows, by renaming $ j \rightarrow b $ on the right,

\begin{equation}
\frac{1}{4} \langle \lambda_a^\dagger \sigma^\alpha \lambda_b \rangle = T_{L,ab}^\alpha \ .
\end{equation}
Analogously for the right transformations:

\begin{equation}
\frac{1}{4} \langle \lambda_a \sigma^\alpha \lambda_b^\dagger \rangle = T_{R,ab}^\alpha \ .
\end{equation}
Written out, this gives the expressions given in Equations (\ref{eqn:leftright}) for the generators.

Now, we discuss the relation to the Higgs doublet.
Defining the doublet as

\begin{equation}
\phi= \begin{pmatrix}
h^2 + i h^1 \\
h^4 - i h^3
\end{pmatrix}
= \begin{pmatrix}
h_u \\ h_d
\end{pmatrix}
\label{eqn:higgsdoublet}
\end{equation}
and its conjugate $ \tilde{\phi} = i \sigma^2 \phi^* $, we find the pseudo-real matrix

\begin{equation}
\Phi = \frac{1}{\sqrt{2}} \lambda^a h^a = \frac{1}{\sqrt{2}} \begin{pmatrix}
h^4 + i h^3 & h^2+ih^1 \\
-h^2 +i h^1 & h^4 -i h^3
\end{pmatrix}
= \frac{1}{\sqrt{2}}(\tilde{\phi},\phi) \ .
\end{equation}
According to the transformation properties (\ref{eqn:chiraltrafo}) and the SM gauging, $ \phi $ and $ \tilde{\phi} $ have hypercharges $ \mp 1/2 $, respectively.
Under the SM subgroup $ SU(2)_L \times U(1)_Y $ we have the decomposition $ \mathbf{4} = (\mathbf{2},\mathbf{2}) \rightarrow \mathbf{2}_{\mathbf{1/2}} $ of the real quadruplet.
This means that the four real components of $ \mathbf{4}=(\mathbf{2},\mathbf{2}) $ form the complex Higgs doublet.

The embedding of the SM fermion doublets into the $ SO(4) $ subgroup of $ SO(6) $ in Section \ref{fermion} differs from that of the Higgs in the aspect that the components are not restricted to be real.
The construction is the same, however, and we obtain a complex $ 2 \times 2 $ matrix, or a complex bidoublet \cite{Wulzer}

\begin{equation}
\Psi = \frac{1}{\sqrt{2}} \lambda^a \psi^a = \frac{1}{\sqrt{2}} \begin{pmatrix}
\psi^4 + i \psi^3 & \psi^2+i\psi^1 \\
-\psi^2 +i \psi^1 & \psi^4 -i \psi^3
\end{pmatrix} \ .
\label{eqn:psicomplex1}
\end{equation}
We give this representation the name $ \mathbf{4}_c = (\mathbf{2},\mathbf{2})_c $.
We can write it as

\begin{equation}
\Psi = (\Psi_- , \Psi_+) = \begin{pmatrix}
\Psi_-^u & \Psi_+^u \\
\Psi_-^d & \Psi_+^d
\end{pmatrix} \ ,
\label{eqn:psicomplex2}
\end{equation}
where the $ \Psi_\pm $ have hypercharge $ \pm 1/2 $.
Since the $ \psi^a $ are complex, there is no such relation between $\Psi_- , \Psi_+ $ as for $ \tilde{\phi},\phi $.
The decomposition thus reads $ \mathbf{4}_c = (\mathbf{2},\mathbf{2})_c \rightarrow \mathbf{2}_{\mathbf{1/2}} \oplus \mathbf{2}_{\mathbf{-1/2}} $.
The $ \mathbf{4}_c $ vector is then, from (\ref{eqn:psicomplex1}) and (\ref{eqn:psicomplex2}), written in terms of the $ \Psi_\pm^{u,d} $,

\begin{equation}
\psi^a = \frac{1}{\sqrt{2}} \ (-i \Psi_+^u -i \Psi_-^d, \ \Psi_+^u - \Psi_-^d, \ i \Psi_+^d -i\Psi_-^u, \ \Psi_+^d + \Psi_-^u)^T \ .
\label{eqn:4cvector}
\end{equation}
This form is used for the quark embedding in Section \ref{fermion}.

\subsection{SO(6) Generators}
\label{SO6}

We take the following basis of generators of $ SO(6) $.

\begin{itemize}

    \item First, the subgroup generators of $ SO(4) \simeq SU(2)_L \times SU(2)_R $:

\begin{equation}
\begin{aligned}
& T_{L,ij}^{a} = -\frac{i}{2} \left[ \frac{1}{2} \epsilon^{abc} (\delta_i^b \delta_J^c - \delta_j^b \delta_i^c) + (\delta_i^a \delta_j^4 - \delta_j^a \delta_i^4) \right] \ , \\
& T_{R,ij}^a = -\frac{i}{2} \left[ \frac{1}{2} \epsilon^{abc} (\delta_i^b \delta_J^c - \delta_j^b \delta_i^c) - (\delta_i^a \delta_j^4 - \delta_j^a \delta_i^4) \right] \ ,
\end{aligned}
\label{eqn:leftright}
\end{equation}
with $ a=1,2,3 $.

    \item Second, the $ SO(5) / SO(4) $ coset generators:

\begin{equation}
T_{5,ij}^a = - \frac{i}{\sqrt{2}} \left[ \delta_i^a \delta_j^5 - \delta_j^a \delta_i^5\right]  \ ,
\end{equation}
with $ a=1,2,3,4 $.

    \item Last, the $ SO(6) / SO(5) $ coset generators:

\begin{equation}
\hat{T}_{ij}^{a} = - \frac{i}{\sqrt{2}} \left[ \delta_i^{a} \delta_j^6 - \delta_j^{a} \delta_i^6 \right]  \ ,
\end{equation}
with $ a=1,2,3,4,5 $.

\end{itemize}
We conventionally decorated also the indices of the cosets with a hat, like $ \hat{a} $, in the main text.
We do without here, for convenience.

The generators are normalized such that $ \langle T^\alpha T^\beta \rangle = \delta^{\alpha \beta} $.
We find the following commutation relations, where Latin indices are now confined to the set $\lbrace 1,2,3 \rbrace$:

\begin{equation}
[T_L^a,T_L^b] = i \epsilon^{abc} T_L^c \ , \ [T_R^a,T_R^b] = i \epsilon^{abc} T_R^c \ , \ [T_L^a,T_R^b] = 0 \ ,
\end{equation}

\begin{equation}
[T_5^a,T_5^b] = \frac{i}{2} \epsilon^{abc}(T_L^c + T_R^c) \ , \ [ T_5^a,T_5^4] =  \frac{i}{2} (T_L^a - T_R^a) \ ,
\end{equation}

\begin{equation}
[\hat{T}^a,\hat{T}^b] = \frac{i}{2} \epsilon^{abc}(T_L^c + T_R^c) \ , \ [ \hat{T}^a,\hat{T}^4] =  \frac{i}{2} (T_L^a - T_R^a) \ ,
\end{equation}

\begin{equation}
[T_{L,R}^a ,T_5^b] = \frac{i}{2} (\epsilon^{abc} T_5^c \pm \delta^{ab} T_5^4) \ , \ [T_{L,R}^a ,T_5^4] = \mp \frac{i}{2} T_5^a \ ,
\end{equation}

\begin{equation}
[T_{L,R}^a ,\hat{T}^b] = \frac{i}{2} (\epsilon^{abc} \hat{T}^c \pm \delta^{ab} \hat{T}^4) \ , \ [T_{L,R}^a ,\hat{T}^4] = \mp \frac{i}{2} \hat{T}^a \ ,
\end{equation}

\begin{equation}
[ \hat{T}^a,\hat{T}^5] = \frac{i}{\sqrt{2}} \ T_5^a \ , \  [T_5^\alpha , \hat{T}^\beta] = \frac{i}{\sqrt{2}} \ \delta^{\alpha\beta} \hat{T}^5 \ ,
\end{equation}

\begin{equation}
[T_{L,R}^a ,\hat{T}^5] = 0 \ .
\end{equation}
The first equation shows that the $ T_{L,R} $ generators are suitably defined to form the $ SU(2)_{L,R} $ sub-algebras.
The last equation shows that $ \hat{T}^5 $ generates a $ U(1) $ symmetry which is not affected by  $ SU(2)_{L,R} $.
It is the generator associated to the Goldstone $ \eta $ in (\ref{eqn:originalGoldstones}).
In other words, it generates the symmetry under which $ \eta $ shifts.
It corresponds to a rotation in the 5-6 plane, $ SO(2)_{5-6} \doteq SO(2)_\eta \simeq U(1)_\eta $.
It is this $ SO(2) $ subgroup of $ SO(6) $ that the embedding of the right-handed quarks must explicitly break in order to give a mass to $ \eta $.
We observe that for $  \epsilon = \pm 1 $, $ \hat{T}^5 \, q_R = \mp \, 1/\sqrt{2} \, q_R $.
Then, the right-handed quarks have a well-defined charge under $ U(1)_\eta $ which is thus unbroken.
The Nambu-Goldstone nature of $ \eta $ is therefore not affected by the SM gauging of $ SU(2)_L \times U(1)_Y $, with $ U(1)_Y $ generated by $ T_R^3 $ (neglecting $ U(1)_X $ in the discussion of the Goldstone sector -- Goldstones are uncharged under $ U(1)_X $), as the name \textit{gauge singlet} for $ \eta $ suggests.
Therefore, for the singlet, no potential and mass are generated by gauge boson loops as we discussed in sections \ref{so6goldstone} and \ref{so6gauge}.

We use the commutation relations to find the Higgs doublet in the $ SO(6)/SO(5) $ coset.
The Higgs doublet comprises four of the five Goldstones parameterizing the coset.
The requirement is that it transforms appropriately as the SM $ SU(2)_L $ doublet.

Defining

\begin{equation}
T_\phi = \begin{pmatrix}
\hat{T}^2 + i \hat{T}^1 \\ \hat{T}^4 -i \hat{T}^3 
\end{pmatrix} \ ,
\end{equation}
we have

\begin{equation}
[T_L^a, T_\phi ] = - \frac{1}{2} \sigma^a T_\phi \ , \ [T_R^3, T_\phi] = -\frac{1}{2} T_\phi \ .
\end{equation}
With this we compute the transformation of $ T_\phi $ under the $ SU(2)_L $.
This is to say a transformation of broken generators under the unbroken group.
Applying the transformation yields

\begin{equation}
\begin{aligned}
e^{-i\alpha_a T_L^a} \ T_\phi \ e^{i \alpha_b T_L^b}& = T_\phi - i \alpha_a [T_L^a, T_\phi] + ... = (1+i \alpha_a \frac{\sigma^a}{2}) T_\phi + ... = e^{i \alpha_a \frac{\sigma^a}{2}} T_\phi \ .
\end{aligned}
\end{equation}
We thus get the sought transformation property.
Therefore,

\begin{equation}
\phi= \begin{pmatrix}
h^2 + i h^1 \\
h^4 - i h^3
\end{pmatrix}
\end{equation}
has the correct quantum numbers to be identified with the Higgs doublet, compare to Equation (\ref{eqn:higgsdoublet}) \cite{carmona}.

\bibliography{Bibliography}

\end{document}